\RequirePackage{rotating}
\RequirePackage{lineno}
\PassOptionsToPackage{usenames,dvipsnames}{xcolour}

\documentclass[a4paper,twocolumn, numberedappendix, twocolappendix,revtex4,iop]{openjournal} 
\usepackage[a4paper,marginpar=10pt,headheight=20pt, right=1cm, top=1.5in, left=1in, bottom=0pt, footskip=0pt, footnotesep=10pt]{geometry}

\pdfoutput=1 
\usepackage{amsmath,amstext}
\usepackage[T1]{fontenc}
\usepackage{apjfonts}
\usepackage{ae,aecompl}
\usepackage[utf8]{inputenc}
\usepackage{natbib}
\usepackage{url}
\usepackage{mdwlist}
\usepackage{listings}
\usepackage{rotating}
\usepackage{multirow}
\usepackage{times}
\usepackage{booktabs}
\usepackage{multirow}
\usepackage{placeins}
\usepackage{fontawesome}

\usepackage[colorlinks,linkcolor=blue,citecolor=blue,urlcolor=blue ]{hyperref}
\usepackage{xurl} 
\usepackage{xspace}
\usepackage{pifont}
\usepackage{orcidlink}
\usepackage{graphicx}
\usepackage{subcaption} 
\usepackage{xcolor}          
\definecolor{orange}{RGB}{255,132,0}
 
\usepackage{tabularx}

\usepackage{ulem}
\usepackage{comment}

\shorttitle{Comparing cosmic shear nulling methods for Stage-IV surveys}
\shortauthors{Robertson \& Hall}

\begin{document}
\title{Comparing cosmic shear nulling methods for Stage-IV surveys}

\author{
    Naomi Clare Robertson$^{1\star}$
    and
    Alex Hall$^{1}$
    }
\thanks{$^\star$ \href{mailto:naomi.robertson@ed.ac.uk}{naomi.robertson@ed.ac.uk}}
\affiliation{
    $^{1}$Institute for Astronomy, University of Edinburgh, Royal Observatory, Blackford Hill, Edinburgh EH9\,3HJ, UK
    }

\begin{abstract}
We present an analysis comparing nulling strategies for reducing the impact of baryon feedback on cosmic shear measurements. We consider three different approaches which aim to `null' the high-$k$ modes using transformations applied to the data vector: the Bernardeau-Nishimichi-Taruya (BNT) transform which operates on the lensing field, a new implementation of an LU factorisation of the discretized Limber integral (LUnul) which operates on the lensing two-point statistics, and finally a method which uses a correlated LSS tracers to suppress contributions from lower redshifts (cross-correlation). We compare these methods to un-nulled (or standard) cosmic shear at the data vector level and assess whether these methods are able to reduce the bias on cosmological constraints using a Fisher forecast. We find that the nulling techniques considered can have a large impact on reducing the bias on $S_8$ and Dark Energy parameters. The cross-correlation method is effective at reducing biases in $S_8$, but requires additional information from galaxy clustering. The LUnul method is the most aggressive of the methods and hence reduces biases most efficiently as $k_{\rm max}$ is increased, although this improvement in accuracy comes at the cost of precision. The BNT approach preserves more information than LUnul, and has a more rigorous theoretical grounding. We demonstrate that all three of these methods are effective at mitigating bias, and can be readily applied in forthcoming lensing analyses.

\end{abstract}

\maketitle

\section{Introduction}
\label{sec:intro}

Cosmic shear measures the clustering of the large-scale-structure (LSS) from the highly non-linear and non-Gaussian sub-Mpc regime, out to much larger linear scales. By measuring the correlation of galaxy shapes at and between different redshifts, the evolution of the LSS can be traced, enabling us to detect the effect of dark energy on the growth of structure. Since gravitational lensing is not sensitive to the dynamical state of the intervening masses, it yields a direct measure of the total mass, which is predominantly dark matter. As such, cosmic shear is now a well established cosmological tool, with stage III surveys providing tight constraints on the matter density parameter $\Omega_{\rm m}$ and the amplitude of matter fluctuations $\sigma_8$ \citep{amon/etal:2022,dalal/etal:2023,wright/etal:2025}. More specifically cosmic shear is most sensitive to the combination of these two parameters given by $S_8=\sigma_8 \sqrt{\Omega_{\rm m}/0.3}$. Photometric galaxy surveys, such as those performed by the Euclid space telescope \citep{euclid/mellier:2024} and Rubin observatory \citep{mandelbauam/etal:2018}, will soon be observing almost the entire observable extragalactic sky, with cosmic shear measurements being a central part of their scientific program. These surveys will therefore benefit from having access to higher statistical power at the larger scales which are easier to model. 
Conversely this means that to gain any further constraining power from weak lensing data sets will require including smaller scales where modelling the non-linear signal is a known challenge \citep[for example,][]{preston/etal:2023}, in particular baryon feedback poses a large uncertainty at these scales. While the constraining power for current surveys has meant that current methods have been sufficient, predictions for upcoming analyses from Euclid and Rubin observatory have forecast severe scale cuts (an $\ell_{\rm max}=1500$ or $\ell_{\rm max}=2000$ depending on the forecast) would be required to recover unbiased cosmological constraints \citep{martinelli/etal:2021,huang/etal:2019}, whilst alternative approaches to modelling these scales with the necessarily larger parameter space leads to a huge decrease in precision \citep{semboloni/etal:2013, huang/etal:2019,spuriomancini/bose:2023,boruah/etal:2024}. 

Baryon feedback generally refers to any process by which the dynamics of baryons present influences that of the wider matter distribution. This includes feedback processes from Active Galactic Nuclei which cause gas to be ejected from massive haloes, and supernovae and star formation processes, which are the dominant effects at $k\gtrsim 10 {\rm Mpc}^{-1}$. 
These processes suppress clustering at small scales and therefore measurements of cosmic shear are impacted. Cosmological hydrodynamical simulations have been able to establish predictions for how baryon feedback processes suppress the matter power spectrum \citep{schaye/etal:2010,vandaalen/etal:2020,schaye/etal:2023,salcido/etal:2023,pakmor/etal:2023,mccarthy/etal:2023,schaller/etal:2025}, however the predicted amplitude of this suppression varies from less than 10\% to more than 30\% around $k=4\, {\rm Mpc}^{-1}$, and can still be up to 10\% at $k=0.3\, {\rm Mpc}^{-1}$, depending on the simulation. Furthermore, simulations with large enough volumes relevant for clustering studies are unable to resolve the small scales required to fully include feedback processes. To get around this, these processes are included via the sub-grid physics, however it has been shown that particular simulation observables are susceptible to the details of these prescriptions (e.g., \citealt{2020MNRAS.492.2285D,2021MNRAS.500.2316C}). Incorrect specification of these effects will bias the resulting matter power spectrum prediction, and therefore the inferred cosmological parameters as well (e.g., \citealt{2018MNRAS.480.2247C, 2020MNRAS.493.1640C}).

Previous analyses have linked the suppression of small scale matter clustering to the baryon fraction of groups and clusters \citep{semboloni/etal:2013,vandaalen/etal:2020,salcido/etal:2023,salcido/etal:2025}. As such, observations that are sensitive to the gas distribution have been used to constrain the impact of baryon feedback or calibrate simulations. X-ray observations, which are sensitive to the hot gas inside galaxies and clusters, have been combined with other probes that measure the total mass, like weak lensing, to estimate the hot gas fraction \citep[for example][]{sun/etal:2009,lovisari/etal:2015,eckert/etal:2016,akino/etal:2022,ferreira/etal:2024,laposta/etal:2024}. Alternatively, measurements of the Sunyaev-Zel'dovich (SZ) effect from CMB experiments have been shown to be able to constrain baryon feedback, either alone \citep{bolliet/etal:2018,reichardt/etal:2021,efstathiou/etal:2025} or in combination with weak lensing \citep{amodeo/etal:2021,troester/etal:2022,pandey/etal:2023,mccarthy/etal:2023,to/etal:2024,bigwood/etal:2024,mccarthy/etal:2024,hadzhiyska/etal:2024,mccarthy/etal:2024}. Data from probes that trace baryonic effects directly will potentially be sufficient for surveys to achieve the accuracy and percent-level precision for which they have been designed, however \citet{wayland/etal:2025} predict that for this to be achievable a CMB-S4-like survey or X-ray data from around 5000 clusters would be required, which are both several years away. 

The contamination of weak lensing measurements from small-scale uncertainties has motivated analysis teams to remove correlations at small angular scales from their data vectors~\citep{2021arXiv210513548K, 2025arXiv251005539E}. This approach has the disadvantage that it does not cleanly remove the small physical (3D) scales where baryon feedback is significant, due to the fact that a given angular scale receives contributions from a wide range of physical scales, a consequence of the broad radial lensing kernel. Weak lensing measures the 2D projection of the matter distribution, meaning large-scale fluctuations from distant lenses contribute to the same angular scales as small-scale fluctuations from nearby lenses. Applying cuts in angular space, e.g., via a maximum harmonic-space multipole $\ell$, is therefore not optimal. 

An alternative approach, and the topic of this paper, is to `null' the contribution from high-$k$ modes by first applying appropriate transformations of the data vector. Many such nulling transformations have been explored in the literature, e.g.,~\cite{hutererandwhite:2005, joachimi/etal:2009, joachimi/etal:2010, bnt:2014, taylor/etal:2018, 2020MNRAS.492.3420B, taylor/etal:2021, touzeau/etal:2025, piccirilli/etal:2025}. Broadly speaking, these methods fall into three categories. Firstly, there are methods that operate on the lensing field (e.g., shear maps) itself. The most prominent method in this category is the Bernardeau-Nishimichi-Taruya (BNT) transform \citep{bnt:2014}, which uses a linear construction of tomographic lensing maps to localise the lensing kernel, allowing a cleaner mapping between angular space and 3D space. Secondly, there are methods that operate on the lensing two-point statistics. This includes the methods discussed in~\cite{hutererandwhite:2005}, and the extensions of the techniques proposed in~\cite{joachimi/etal:2009}. Thirdly, there are methods that use complementary tracers of the LSS to remove contributions from specific redshifts. These have been mostly developed in the context of CMB lensing, for example~\cite{2021PhRvD.103j3538M, qu/etal:2023}, in order to decompose the lensing signal into contributions from specific redshift slices. In the Limber approximation of the weak lensing spectrum at a given angular scale, cutting in redshift is equivalent to cutting in 3D scale, such that these methods can be generalised to cosmic shear in the presence of a complementary foreground galaxy clustering sample, as we show in this paper. The presence of such samples is becoming commonplace as part of `3$\times$2-pt' analyses (see~\citealt{2025arXiv250919514L} for a recent proposal to use this method in combination with Fast Radio Bursts). Whilst this list of methods is not exhaustive, it broadly captures the most popular techniques explored by the community. In particular, we do not consider recent innovations such as those of \citet{2021MNRAS.507.5592M, maraio/etal:2025}, who include a theoretical covariance to capture the uncertainty of modelling at small scales, or \citet{derose/etal:2025}, who take an effective field theory inspired approach by including lensing counter terms to isolate the large scale cosmological signal from any small scale physics. 

In this work, we conduct an assessment of the performance of these three approaches in the context of the forthcoming Stage-IV cosmic shear surveys that aim to constrain Dark Energy physics at the percent level. As part of this exercise, we develop the `$k$-cut' method presented in~\cite{hutererandwhite:2005}, which to our knowledge has not been considered in the literature since it was first proposed. We also adapt the cross-correlation method from the CMB lensing literature and apply it to cosmic shear in combination with photometric galaxy clustering. 

The overall aim of this paper is to inform the analysis of Stage-IV surveys by proposing improvements to the standard scale cut methodology through the use of nulling techniques, and therefore to make cosmological inference unbiased by uncertainties in non-linear structure growth while retaining as much information as possible.

This paper is organised as follows: in section~\ref{sec:theory} we briefly introduce the cosmic shear formalism and the nulling methods considered in section~\ref{sec:nulling}, in section~\ref{sec:mock} we describe the mock data vector we use to compare these methods, and show the impact of nulling methods at a data vector level in section~\ref{sec:null_impact} and on cosmological constraints using a Fisher analysis in section~\ref{sec:fisher}. Our conslusions are presented in section~\ref{sec:discussion}.

\section{Cosmic Shear Theory}
\label{sec:theory}
In this section we briefly introduce some cosmic shear formalism (see~\citealt{kilbinger/etal:2017} for further details). For a pair of tomographic bins, $i$ and $j$, an accurate expression for the cosmic shear power spectrum, valid in the Limber approximation, is given by
\begin{equation}
    C_{\ell,\gamma \gamma}^{ij} = \int_0^{\chi_{{\rm max}}} \mathrm{d} \chi \, \frac{q^{i}(\chi) q^{j} (\chi)}{\chi^2} \, P_{\rm mm}\left( k=\frac{\ell + 1/2}{\chi} ,\, z(\chi) \right) \, ,
    \label{eq:limber}
\end{equation}
for a (non-linear) matter power spectrum $P_{\rm mm}(k)$ where $k$ is the 3D wavenumber, $\ell$ is the 2D multipole, $\chi$ is the comoving distance with maximum $\chi_{{\rm max}}$ in the survey, and $q^i(\chi)$ is the weak lensing window function for bin $i$, given by
\begin{equation}
    q^i (\chi) = \frac{3 H_0^2 \Omega_{\rm m}}{2 c^2}  \, \frac{\chi}{a(\chi)} \, g^i(\chi) \, ,
\end{equation}
where the kernel for tomographic bin $i$ is given by 
\begin{equation}
    g^i (\chi) = \int_{\chi}^{\chi_{\rm max}} \mathrm{d} \chi' \, n_{\rm s}^i(\chi')\, \frac{\chi' - \chi}{\chi'} \, ,
\end{equation}
for a source redshift distribution $n_{\rm s}^i(z) =  n_{\rm s}^i(\chi)  \,d\chi/dz$. 
From Equation~\ref{eq:limber} we can see that for a given value of $\ell$ a range of $k$ modes at different values of $\chi$ (or redshifts $z$) contribute to the cosmic shear power spectrum.
  
\section{Nulling Methods}
\label{sec:nulling}
In this section we describe each of the nulling methods considered in this analysis.

\subsection{LU nulling}

We describe here a new implementation of a method first proposed in~\cite{hutererandwhite:2005}, which to our knowledge has not been since considered in the literature. We will hereafter refer to this method as `LUnul', and provide a public implementation\footnote{\url{https://github.com/ahallcosmo/lunul}}.

The starting point is to first recast the Limber integral, Equation~\eqref{eq:limber}, as an integral over $\log k$. Then, we approximate this integral as a discrete summation by defining $N_k$ nodes at which the integrand is evaluated. For simplicity, we will assume that $N_k$ is independent of the tomographic bin pair. Next, we recast this summation as a matrix multiplication. By vectorizing over the tomographic bin indices, we can re-write Equation~\eqref{eq:limber} as
\begin{equation}
    \mathbf{c}_\ell = \mathbf{P}_\ell \mathbf{1}
    \label{eq:matlimb}
\end{equation}
where $\mathbf{c}_\ell$ is the vector of angular power spectra for each tomographic bin pair at a given $\ell$; if there are $N$ tomographic bins, this vector has length $N_p =N(N+1)/2$. The matrix $\mathbf{P}_\ell$ has dimensions $N_p \times N_k$, and $\mathbf{1}$ is a vector of ones with length $N_k$. This matrix formulation of the Limber integral has some computational advantages, with the potential to reduce computational time, however the drawback that $N_k$ is required to be the same for every bin pair.

For each $\ell$, we then perform an LU decomposition of $\mathbf{P}_\ell$. Since $N_p < N_k$, this results in a square $N_p \times N_p$ lower triangular matrix $\mathbf{L}_\ell$ with unit diagonal, and a $N_p \times N_k$ upper triangular matrix $\mathbf{U}_\ell$, such that $\mathbf{P}_\ell = \mathbf{L}_\ell \mathbf{U}_\ell$. 

Since $\mathbf{L}_\ell$ is square, triangular, and has unit diagonal, it is invertible. Applying the inverse to Equation~\eqref{eq:matlimb} gives
\begin{equation}
    \mathbf{L}_\ell^{-1} \mathbf{c}_\ell = \mathbf{U}_{\ell} \mathbf{1} \, .
    \label{eq:LU}
\end{equation}
For a given fiducial model, $\mathbf{L}_\ell^{-1}$ is readily computable, and can be applied to estimates of the power spectra. If the fiducial model is correct, the right-hand side of Equation~\eqref{eq:LU} is the expectation value of this transformed set of spectra. 

Consider now which $k$-modes contribute to each row of Equation~\eqref{eq:LU}, i.e.~to each of the transformed power spectra. Since $\mathbf{U}_\ell$ is upper triangular, the first row receives no contribution from the first column of $\mathbf{U}_\ell$. Since the column space of $\mathbf{U}_\ell$ represents the nodes of the summation over $\log k$ in the Limber integral, we can ensure that the highest-$k$ node does not contribute to the first row of  $\mathbf{L}_\ell^{-1} \mathbf{c}_\ell$ if we order the summation in order of decreasing $k$. Likewise, the second row of $\mathbf{L}_\ell^{-1} \mathbf{c}_\ell$ receives no contribution from the first two columns of $\mathbf{U}_\ell$. Moving down row by row, we will either reach a row where no columns having $k < k_{\rm max}$ contribute, or we will exhaust all of the $N_p$ rows available. 

Our strategy is therefore to define a $k_{\rm max}$ and, for each $\ell$, discard rows from $\mathbf{L}_\ell^{-1} \mathbf{c}_\ell$ that have contribution from $k > k_{\rm max}$. If no such rows are available, we discard all of the rows, which is equivalent to throwing out the whole data vector for that $\ell$. Since higher $k$ contribute progressively more to higher $\ell$ in a smooth fashion, this procedure results in a softer form of the standard scale cut method. Effectively this method is not removing individual tomographic spectra, but instead forming linear combinations of them that cancel specific parts of the Limber integrand. Since high-$k$ power mostly comes from low-redshift structure, where all lensing kernels look similar, the LU decomposition results in the common high-$k$ part being cancelled out, while the differences at lower-$k$ (which are dependent on the redshift/bin pair) are retained.

There is some freedom in the implementation of the LUnul method. Firstly, there is sensitivity to the way in which the Limber integral is discretized, especially because only the first $N_p$ modes can formally be discarded and typically $N_p \ll N_k$. We found a linear spacing in $\log k$ works well, having also experimented with Monte Carlo sampling of the Limber integrand and the list of nodes points returned by the \texttt{scipy.integrate.quad} routine for a fixed choice of bin pair. These latter two choices of nodes can work well for bin pairs close to the fiducial bin pair that generated them, but become suboptimal for other bin pairs (the disadvantage of enforcing a constant $k$-sampling for every bin pair, as referenced earlier). We also introduce a tolerance parameter, which allows rows of $\mathbf{U}_\ell$ to be retained if modes having $k>k_{\rm max}$ contribute less than a fraction \texttt{tol} to the total over all $k$. The results in this paper set  \texttt{tol=0.01}, which gives excellent results. We considered a range a different values for the tolerance parameter and found the same trends. We also have the freedom to choose the number of nodes, $N_k$, used to discretize the integral. In general, this should be large enough to attain accuracy in the power spectrum within some chosen tolerance, but small enough that the method retains some of its unique features; if $N_k$ is too large, then the first $N_p$ rows of the transformed data vector will only have the highest $k$ nodes nulled out, and the method will essentially be equivalent to discarding $\ell$-modes that have more than a fraction \texttt{tol} contribution from $k>k_{\rm max}$. We have found $N_k=\mathcal{O}(100)$ works well.

\subsection{BNT}
The BNT transform~\citep{bnt:2014} is a linear transform which reorganises the cosmic shear signal by separating the contribution from foreground structure at different redshifts. The transformed lensing kernels are narrower and as such the resulting power spectra has a closer mapping between $k$ and $\ell$. All information is preserved since the BNT transform is linear and invertible. This method benefits from a large number of tomographic bins and therefore has had limited impact of current Stage III surveys \citep{taylor/etal:2021} but has been shown to be able to significantly improve constraining power for a Euclid-like survey \citep{gu/etal:2025}.

The BNT transform reorganises the cosmic shear data vector $C_{\ell,\gamma \gamma}^{ij}$ to a new data vector $\hat{C}_{\ell,\gamma \gamma}^{ab}$, where $a$ and $b$ refer to the new lensing window functions given by
\begin{equation}
    \hat{q}^a(\chi) = \sum_{i=1}^{N} f^a_i q^i(\chi) \, ,
\end{equation}
where $N$ is the number of tomographic bins and $f_i^a$ are the BNT transform coefficients. The transformation only acts on the tomographic bin space (mapping indices $i$ to transformed indices $a$), leaving the angular space unchanged.

The coefficients $f_i^a$ are chosen as the solutions to the algebraic equations
\begin{align}
    \sum_{i=a-2}^a f_i^a \int d\chi \, n_{\rm s}^i(\chi)=0 \, , \\
    \sum_{i=a-2}^a f_i^a \int d\chi \frac{n_{\rm s}^i(\chi)}{\chi}=0 \, .
\end{align}
With these choices of coefficients, the BNT transform aims to localise the lensing kernel such that cross-correlations between tomographic bins that are not directly next to one another are suppressed, with
\begin{equation}
    \hat{q}^a(\chi)\times\hat{q}^b(\chi)=0 \,\,\, {\rm{for}} \,\,\, |a-b|\geq 2 \, .
\end{equation}
The BNT transformed power spectrum is given in terms of the untransformed power spectra by
\begin{align}
\hat{C}_{\ell,\gamma \gamma}^{ab} & = \int d\chi \frac{\hat{q}^a(\chi)\hat{q}^b(\chi)}{\chi^2} P_{\rm mm}\left( k=\frac{\ell + 1/2}{\chi} , z(\chi) \right) \, \nonumber \\
& = \sum_{i,j}f_i^a f_j^b C_{\ell,\gamma \gamma}^{ij} \, .
\end{align}

The advantage of the BNT transform is that it operates directly on the shear signal, and hence can be useful for any statistics based on the shear signal, not just the two-point statistics. It should be noted, however, that simply re-weighting the maps is not sufficient as it couples the noise, which is problematic for some high-order statistics. Methods have been developed to account for this, by applying the BNT method, smoothing the maps and then transforming back to the original basis as implemented in \citet{2025arXiv251004953E}. Another advantage is that the transformation is linear and invertible, and hence lossless.

After transformation, the mapping between angular space and 3D physical space is much cleaner in the power spectrum. Subsequent cuts in $\ell$ on the transformed power spectra can hence provide a cleaner separation of non-linear scales~\citep{taylor/etal:2018, 2020PhRvD.102h3535D, taylor/etal:2021}. The performance of this method is most effective when many tomographic bins are available. An example of the impact of the BNT transformation on the lensing kernels is described in Section~\ref{sec:mock}.

\subsection{Cross-correlation}
CMB studies have used cross-correlations with foreground galaxy samples to remove the lensing contribution from the matter distribution at the redshifts of these galaxies (e.g., \citealt{qu/etal:2023}). The aim in those analyses was to isolate lensing due to structure at a higher redshift, but its application to cosmic shear holds potential benefits by removing the low redshift contribution. At a fixed angular scale, it is the baryon suppression at low redshifts (equivalent to small physical scales) that contaminates the cosmic shear signal from higher redshift (equivalent to larger, more linear scales). Tracers at lower redshift can hence be used to partially remove the contribution to the shear signal from these low-redshift, small-scale structures.

Assuming we have access to maps of galaxy number density in $N_\delta$ tomographic bins, with fluctuations denoted by $\delta$, we define a de-correlated shear field by
\begin{equation}
    \tilde{\gamma}_{\ell m}^i = \gamma_{\ell m}^i - \sum_{rs=1}^{N_\delta} C_{\ell,\gamma \delta}^{ir} \, \mathbf{C}_{\ell,\delta \delta}^{-1,rs} \, \delta^s_{\ell m}\, ,
    \label{eq:xcorr}
\end{equation}
where the auto-correlation of the galaxy density is given by $\mathbf{C}_{\ell,\delta \delta}$, which is a matrix in the space of tomographic bins of the clustering sample, and the cross-correlation between the two fields is given by $C_{\ell,\gamma \delta}^{ir}$. Note that in Equation~\eqref{eq:xcorr} we have chosen to represent the transformation in harmonic multipole space, but we could just have easily used configuration (i.e.~angular) space as the operation is purely on the tomographic bin indices.

When the cross-correlation and auto-spectra appearing in Equation~\eqref{eq:xcorr} are their true values, it may be easily verified that the quantity $\tilde{\gamma}_{\ell m}^i$ is uncorrelated with all of the galaxy density maps. For practical applications, we can use model predictions for these quantities computed at fiducial values of the cosmological (and galaxy bias) parameters. If these are close to the true values, the transformed field will be approximately uncorrelated with the foreground clustering maps. Note that the auto-spectrum of density fluctuations, $C_{\ell,\delta\delta}$, should include the contribution from shot noise. Here we have assumed a linear galaxy bias for the clustering part, which we marginalise over in the Fisher forecast.

Assuming for now that the power spectra in Equation~\eqref{eq:xcorr} take their true values, the power spectrum of the transformed shear field is given by
\begin{equation}
    \mathbf{C}_{\ell,\tilde{\gamma}\tilde{\gamma}} = \mathbf{C}_{\ell,\gamma \gamma} - \mathbf{C}_{\ell,\gamma \delta} \, \mathbf{C}_{\ell,\delta \delta}^{-1}  \, \mathbf{C}_{\ell,\delta \gamma} \, .
\end{equation}
where $\mathbf{C}_{\ell,\delta \delta}$ is the matrix of power spectra between clustering bins and $\mathbf{C}_{\ell,\gamma \delta}$ is the matrix of cross-correlations between source bins and foreground clustering bins. In the high signal-to-noise regime where shot noise can be neglected, this expression is independent of the linear galaxy bias of the clustering sample. In fact, it is easy to show from Equation~\eqref{eq:xcorr} that errors between the bias parameters assumed in the fiducial model and those in the real sample enter $\mathbf{C}_{\ell,\tilde{\gamma}\tilde{\gamma}}$ quadratically, such that the Fisher matrix of the cosmological parameters is not degraded by lack of knowledge of the linear galaxy bias. This is supported by the power spectrum derivatives with respective to the galaxy bias parameters, which are close to zero, and also by the fact that parameters constraints are only slightly changed when we marginalise over these parameters compared with them being fixed. Note that we fix the cross-correlation and auto-correlation to fiducial values when we decorrelate the shear field from the density field, and include the cosmological parameter dependence in the cross-correlation between the measured density field and the measured shear field.

The construction presented here provides a simple way of damping the contribution to the lensing signal from low-redshift structure. Clearly, for this to work, we need a set of foreground clustering samples at lower redshift than the sources, ideally down to low redshifts where the lensing kernels still have significant support. Such a scenario is provided by samples designed for 3$\times$2-pt analyses, where foreground lens samples aid in mitigating the impact of galaxy intrinsic alignments, photometric redshift uncertainties, and shear measurement biases~\citep{2020A&A...643A..70T, desy3/etal:2022, 2025OJAp....8E..24B}. Indeed, given that the transformation in Equation~\eqref{eq:xcorr} is a linear combination of shear and clustering, one might be wondering if the mitigation we are proposing here is entirely captured by a 3$\times$2-pt analysis. In fact, this is not the case; we do not include the auto-spectra of the clustering sample in our data vector, which would amount to `adding back in' information from the structures we are attempting to remove. Instead, we are only using the clustering sample to `clean' the shear sample. The transformed shear signal, $\tilde{\gamma}$, provides an alternative to the combined shear and clustering signal, and its power spectrum $\mathbf{C}_{\ell,\tilde{\gamma}\tilde{\gamma}}$ an interesting alternative to the full 3$\times$2-pt data vector. 

Note that after performing the transformation Equation~\eqref{eq:xcorr}, we are free to perform additional scale cuts to further suppress the contribution from small spatial scales.

\section{Mock Data}
\label{sec:mock}
\begin{figure}
    \centering
    \includegraphics[width=\columnwidth]{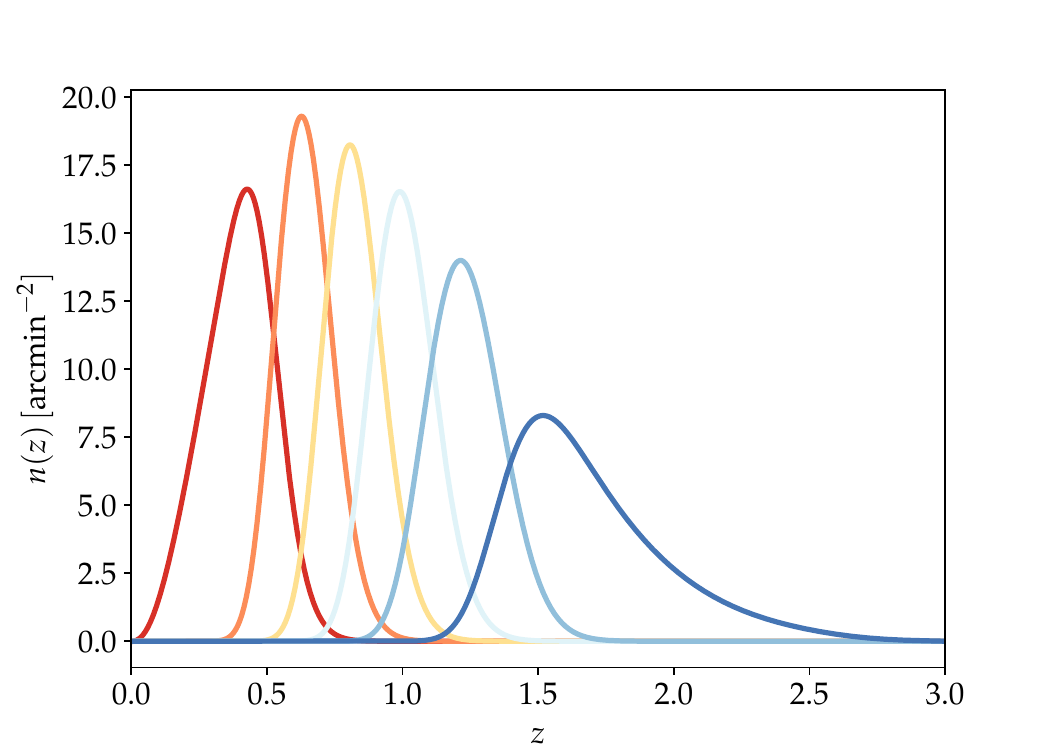}
    \caption{Redshift distributions for six tomographic bins, which are used to produce mock power spectra. For the case of the cross-correlation method we take the two lowest redshift bins as our lens sample and the three highest redshift bins as our source sample, ignoring the third redshift bins to minimise overlap between the lens and source sample.}
    \label{fig:nz}
\end{figure}

In order to test our nulling routines, we create mock lensing power spectra to serve as a fiducial data vector. We consider a Euclid-like survey setup~\citep{euclid/mellier:2024}, with six tomographic source bins shown in Figure~\ref{fig:nz}. Theory predictions are made using the Core Cosmology Library \citep{CCL}\footnote{https://github.com/LSSTDESC/CCL}. We assume a galaxy number density of 30 galaxies$/{\rm arcmin}^2$, an intrinsic ellipticity dispersion $\sigma_e=0.26$ and a sky area of 15,000 ${\rm deg}^2$. For parameter inference, we assume a Gaussian covariance matrix for the power spectra with a simple $f_{{\rm sky}}$ scaling to account for the finite sky area.\footnote{We note that including the super sample covariance reduces information for $\ell >1000$ and therefore adding this term will impact the results.} We assume a $w_0w_{\rm a}$CDM cosmological model with $\{\Omega_{\rm m}=0.3,\, h=0.715,\, \Omega_{\rm b}h^2=0.02,\, S_8=0.8,\, n_s{\rm}=0.97, \,w_0=-1,w_{\rm a}=0  \}$. 

\begin{figure}
    \centering
    \includegraphics[width=\columnwidth]{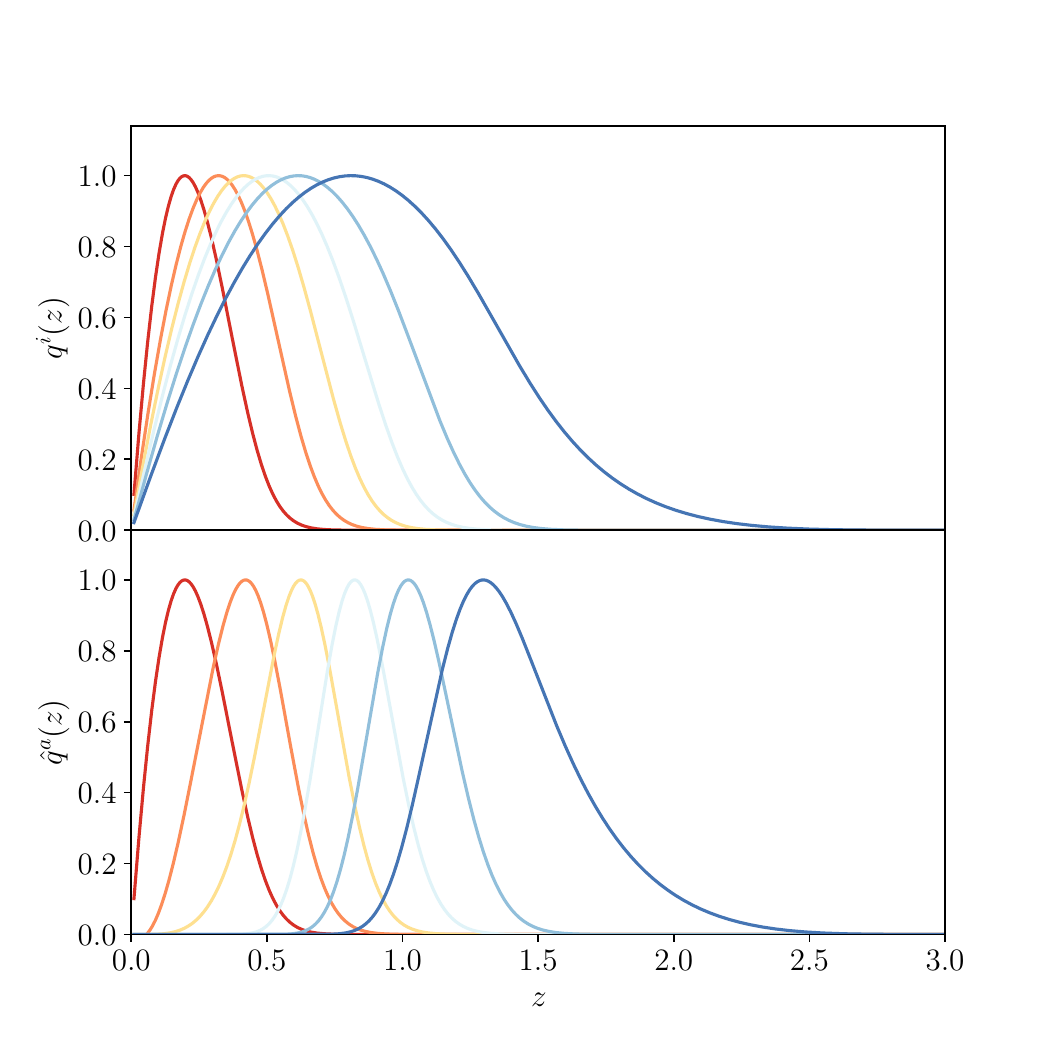}
    \caption{The top panel shows the lensing kernels, $q^i(z)$ for the six tomographic bins shown in Figure~\ref{fig:nz}, and the lower panel shows the BNT transformed lensing kernels, $\hat{q}^a(z)$. These distributions have been normalised to have the same peak value.}
    \label{fig:kernel}
\end{figure}

The cross-correlation method relies on a low-redshift lens sample used to `clean' a higher redshift source sample. When comparing the performance of this method against other methods that use only the source sample, we must be careful to account for the additional information brought by the lens density field. To perform as close an `apples-to-apples' comparison as possible, we fix the total (i.e., lens + source) sample and assume that the lowest two source redshift bins can be repurposed as lens bins. The third source bin has strong overlap with neighbouring bins and is not expected to benefit significantly from the cross-correlation nulling, so we discard it. This configuration of two low-redshift lens bins and three high-redshift source bins should be roughly optimal for this form of nulling, but we caution against over-interpretation of the comparison with the other nulling methods. As the cross-correlation method has not been considered in the context of 3x2pt samples before, we consider this setup sufficient as a first test of its performance. The lensing kernels shown in Figure~\ref{fig:kernel}, for the standard case in the top panel and the BNT-transformed case in the lower panel.

\begin{figure}
    \centering
    \includegraphics[width=\columnwidth]{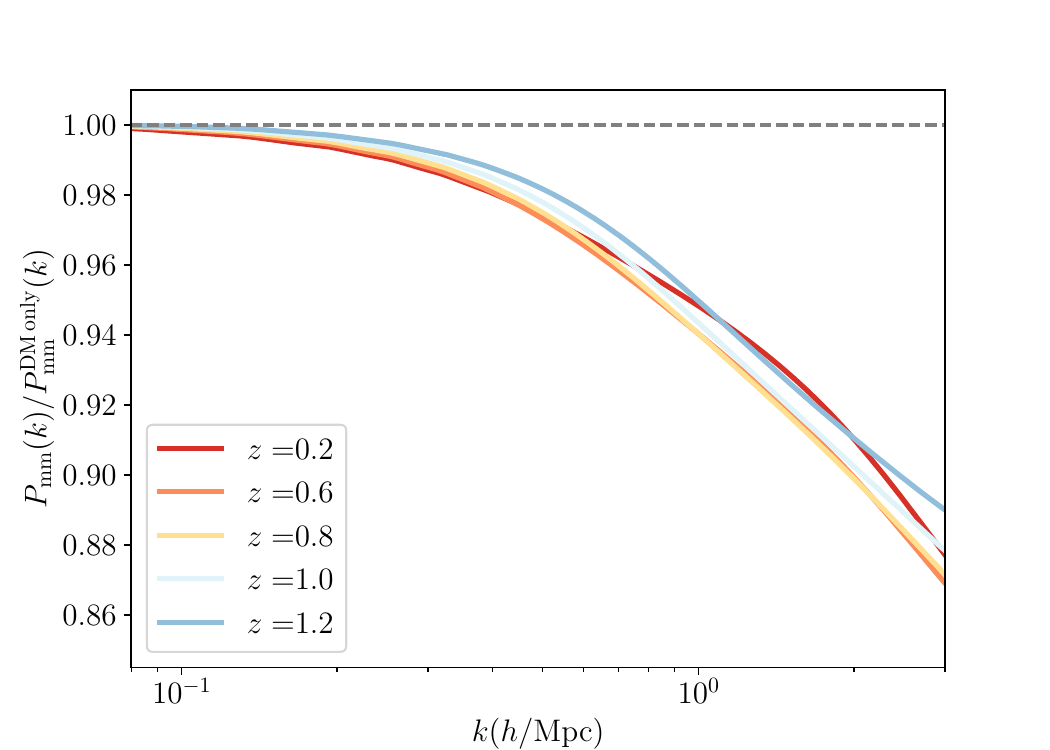}
    \caption{Suppression of the matter power spectrum due to baryon feedback ($\Theta_{\rm AGN}=7.8$) for a range of redshift values probed by the mock cosmic shear data in this analysis. Previous studies have shown that while suppression of the matter power spectrum due to baryons increases with decreasing redshift, it only evolves weakly for $z \lesssim 1$, as shown here \citep{vandaalan/etal:2011,mccarthy/etal:2017}. The evolution is non-monotonic with redshift due to different effects being more or less dominant at different times, for example gas cooling versus feedback processes \citep{chisari/etal:2018}.}
    \label{fig:Pmm}
\end{figure}

Baryon feedback is included using HMCode2020~\citep{mead/etal:2021} via the parameter $\Theta_{\rm AGN}$ which modulates the strength of the baryon feedback suppression. This is not a physical parameter but is related to an AGN heating parameter in the BAHAMAS simulations \citep{mccarthy/etal:2017}, defined as as $\Theta_{\rm AGN}=\log_{10}(\Delta T_{\rm heat}/{\rm K})$, which HMCode2020 was designed to be able to recover. We use $\Theta_{\rm AGN}=7.8$ which matches the fiducial BAHAMAS simulation -- the ratio of the matter power spectrum with and without baryon feedback is shown in Figure~\ref{fig:Pmm} for a range of redshift values.

The ratio of the cosmic shear spectra with and without the suppression due to baryon feedback is shown in Figure~\ref{fig:baryon_sup} (orange curves). The suppression reaches the 10\% level at $\ell \sim 1000$ for low redshift source bins, and $\ell \sim 4000$ for high redshift source bins, reflecting the predominance of smaller spatial scales in the lensing of lower redshift galaxies.

\section{Impact of nulling strategies on the cosmic shear power spectra}
\label{sec:null_impact}
\begin{figure*}
    \centering
    \includegraphics[width=\textwidth]{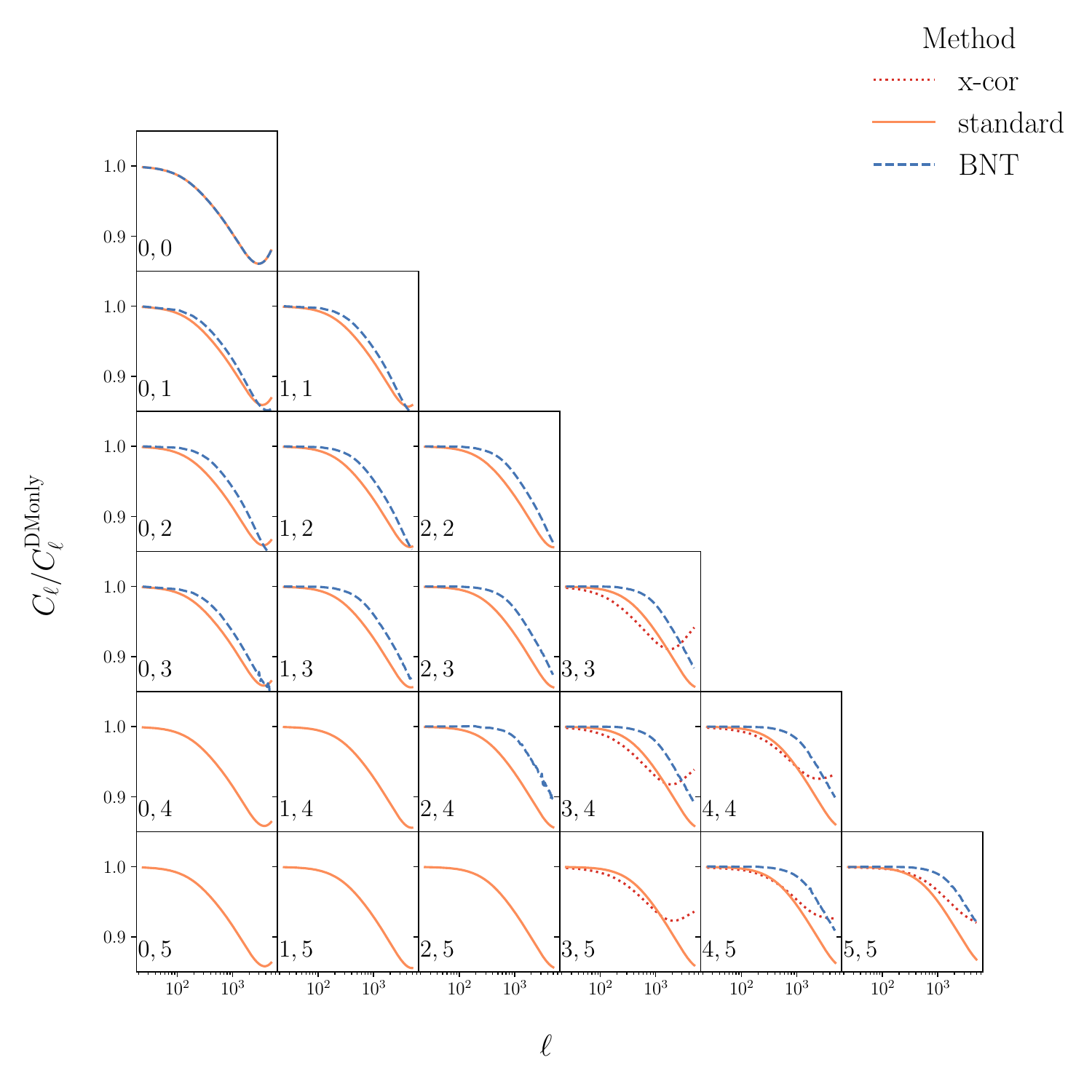}
    \caption{Impact of baryon feedback, assuming $\Theta_{\rm AGN}=7.8$, on the cosmic shear power spectra for the `standard' case (orange), the BNT transformed case (blue) and the cross-correlation case (red). In all but the (0,0) bin pair, the BNT transformed spectra is less impacted by baryons, particularly considering higher redshift bin pairs. The cross-correlation-modified spectra are least impacted by baryons for the highest redshift bin pairs where the separation between lenses and sources is greatest. When there is more overlap between the bins the suppression due to baryons is increased at intermediate $\ell$ compared to the standard case.}
    \label{fig:baryon_sup}
\end{figure*}

\begin{figure*}
    \centering
    \begin{tabular}{c c}
    {\bf Fiducial} & {\bf LUnul Transformed} \\
    \includegraphics[width=0.45\textwidth]{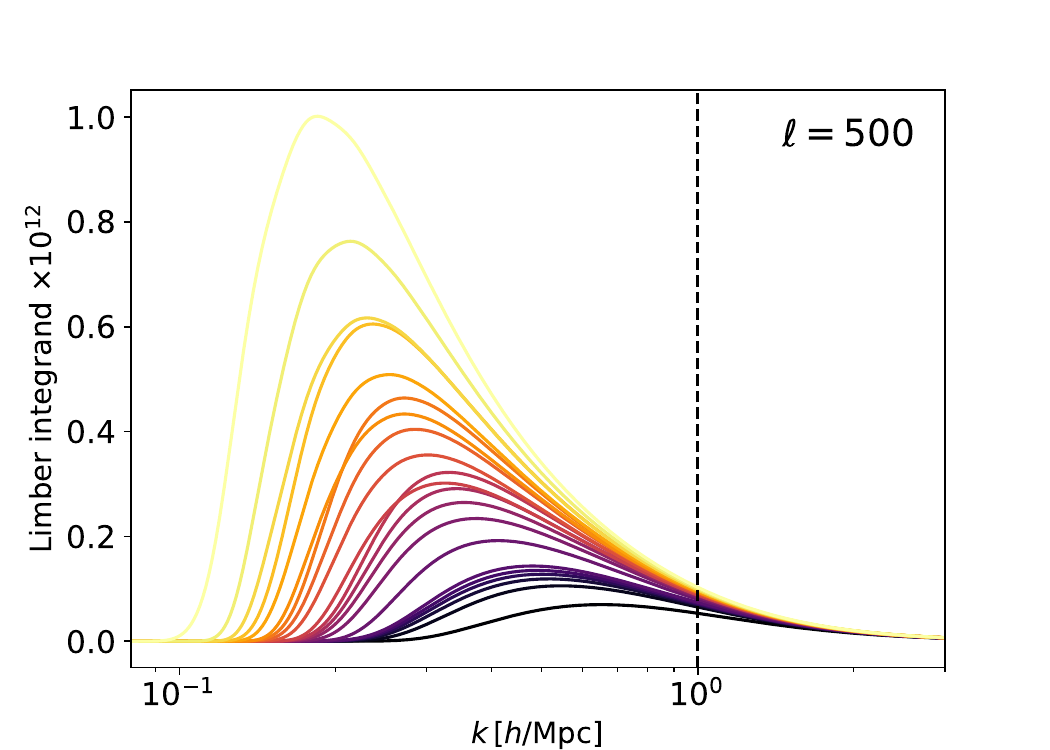} & \includegraphics[width=0.45\textwidth]{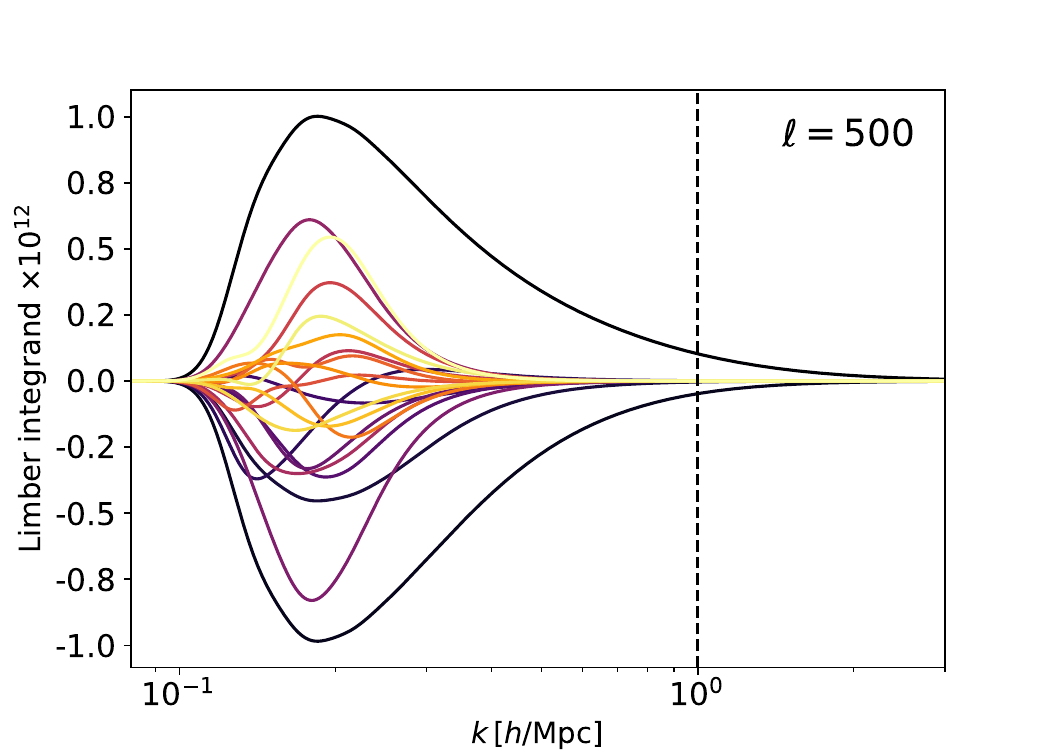} \\
    \includegraphics[width=0.45\textwidth]{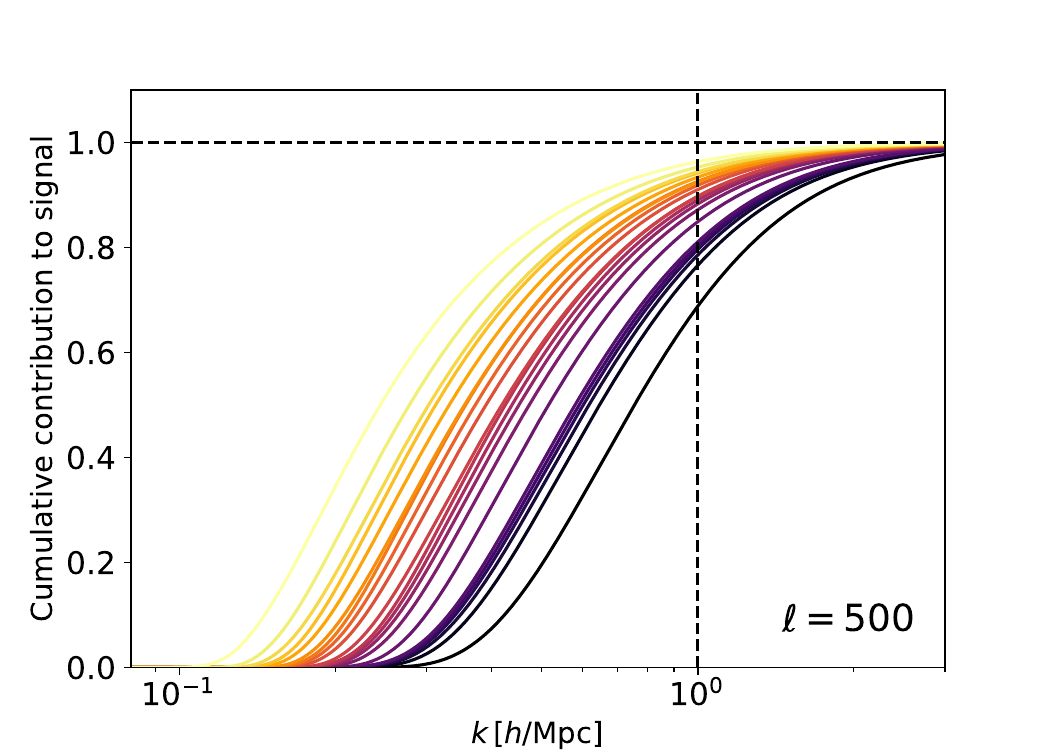} & \includegraphics[width=0.45\textwidth]{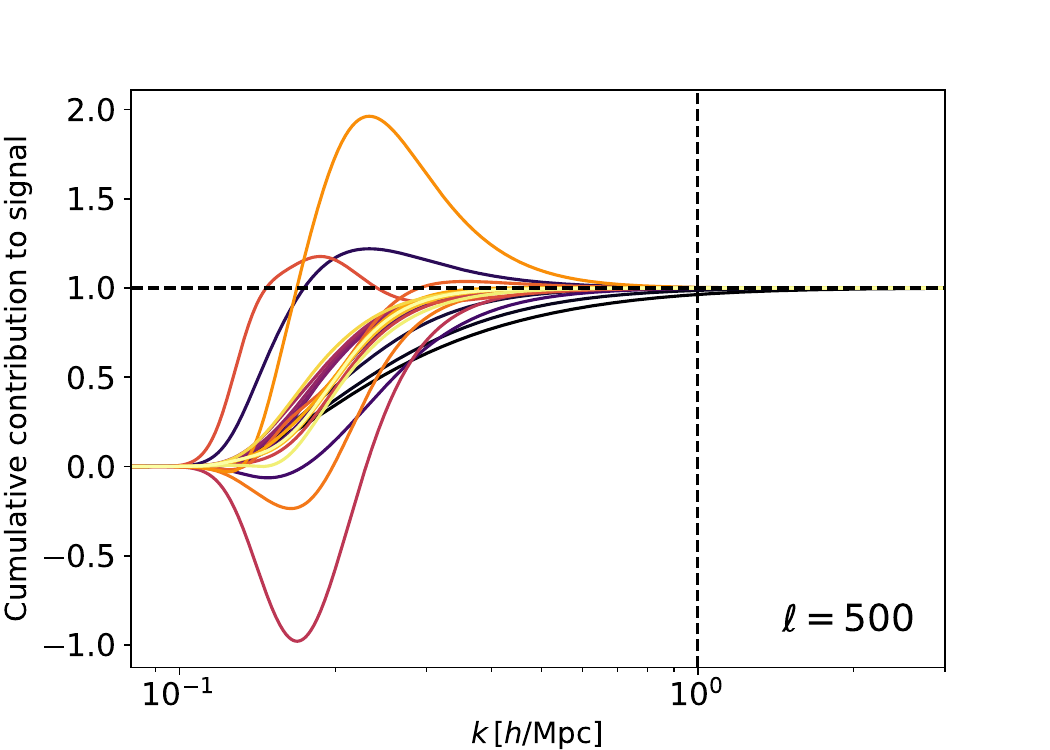}
    \end{tabular}
    \caption{Performance of the LUnul method for $\ell$=500. {\it Upper left panel}: the Limber integrand for each bin pair in our fiducial setup. Different colours refer to different bin pairs, with darker colours corresponding to closer source bins. {\it Lower left panel}: cumulative contribution across $k$, normalized to unity. {\it Upper right panel}: Limber integrand after applying the LUnul transform. Colours now refer to elements of the transformed bin pair space. {\it Lower right panel}: cumulative contribution across $k$, normalized to unity. The vertical dashed line indicates $k= 1 \, h/{\rm Mpc}$. We have chosen to discretize the integral with $N_k=100$ points linearly spaced in $\log k$, and impose a tolerance \texttt{tol=0.01} to small-scale power, as described in the text.}
    \label{fig:LUnulplots}
\end{figure*}

The goal of the nulling methods we are considering is to mitigate the suppression on the cosmic shear signal due to baryon feedback, shown in Figure~\ref{fig:baryon_sup} for the BNT transformed (blue-dashed lines) and the de-correlated (red-dotted lines) cases. The BNT transformed case is only shown for the auto-correlations and close pairs of tomographic bins as the well separated bins have close to zero correlation by design. The `x-cor' case is shown for the highest three redshift bins as the lowest two redshift bins are used as the lens sample and the third bin is neglected to reduce overlap between lens and source galaxies \footnote{We tested this choice by checking the impact on cosmological parameter constraints when including the third redshift bin shown in~\ref{fig:nz} in the lens sample and found the constraint on $S_8$ to be biased by at least an additional $0.5 \sigma$ compared to the two lens bin case.}. The impact for the LUnul method is not shown here as the LUnul method operates in a transformed space (as described in Equation \ref{eq:LU}), with elements that are no longer indexed by the bin pair indices shown in Figure \ref{fig:baryon_sup}. Instead, we plot the sensitivity to high $k$ effects in Figure \ref{fig:LUnulplots} for the LUnul method.

\begin{figure*}
    \centering
    \includegraphics[width=\textwidth]{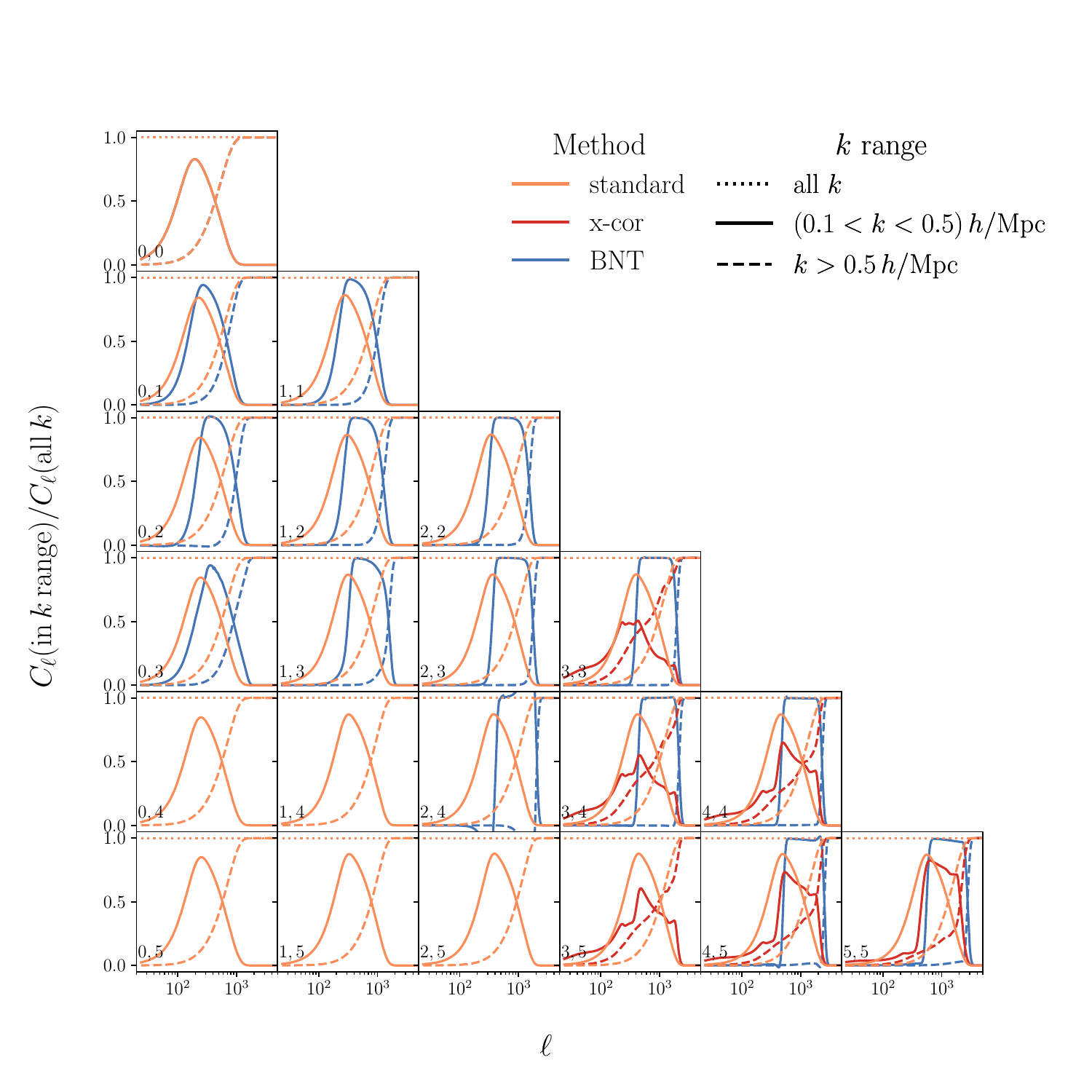}
    \caption{Ratio of power spectra for each pair of tomographic bins, showing the contribution from a given range in $k$ to the overall power spectrum: the dashed line corresponds to the smallest scales with contributions from $k>0.5h/{\rm Mpc}$, the solid line corresponds to slightly larger scales $(0.1<k<0.5)h/{\rm Mpc}$ and the dotted line shows the full spectra including all $k$. The `standard' case (orange) shows small scales contribute to even the smallest $\ell$-modes, whilst the BNT-transformed spectra (blue) have a more direct mapping between $k$ and $\ell$, especially for the higher redshift bins. The cross-correlation-modified spectra (red, highest three source bins only) exhibit a damping of the contribution from small scales relative to the `standard' cosmic shear spectra; this is most significant in the highest redshift bin which has the largest separation to the lens sample. In the cross-correlation case there are `wiggles' visible in the range $(0.1 < k < 0.5)h/{\rm Mpc}$ (red solid line) which come from the first lens bin and is due to the presence of the BAO in the clustering signal.}
    \label{fig:krange-ratio}
\end{figure*}

Figure~\ref{fig:krange-ratio} shows how different ranges of $k$ contributes to the cosmic shear power spectra at each $\ell$ -- this is shown for the standard, BNT and cross-correlation cases in orange, blue and red. In the standard case, the smallest scales $(k>0.5h/{\rm Mpc})$ dominate at $\ell > 1000$, and at low redshift these scales still contribute up to 10\% of the signal at $\ell \sim 500$. For reference this figure is also shown in the appendix, Figure~\ref{fig:krange}, as $C_{\ell}$ rather than as a ratio. 

Under the BNT approach, scales $k>0.5h/{\rm Mpc}$ are limited to $\ell>2000$ for the higher tomographic bins, with their contribution dropping sharply to be minimal for $\ell<2000$. The $\ell$ range widens for the lower tomographic bin pair, but the relationship between $\ell$ and $k$ is consistently more direct than the standard case. Considering the impact this change in $k$-contribution has on the suppression due to baryons (shown in Figure~\ref{fig:baryon_sup}), 
the lensing kernel for the lowest redshift bin in the standard and BNT cases are the same and therefore there is no change in the level of suppression, however when moving to higher redshift bins the impact of baryons is pushed to higher values of $\ell$. For a 5\% suppression in $C_{\ell}$ this corresponds to an $\ell \sim 3000$ compared to $\ell \sim 1000$ without using the BNT transform. 

For the cross-correlation method, we see a similar effect for the highest redshift bin, with a sharp drop in the $(k>0.5h/{\rm Mpc})$ scales contributing to $\ell<2000$. For the two lower redshift source bins, we see that the intermediate $k$-range $(0.1<k<0.5)h/{\rm Mpc})$ contribute to a wide range of $\ell$-values, however this contribution is suppressed to $\sim50$\% compared to the un-nulled case. The prominent `wiggles' visible in Figure~\ref{fig:krange-ratio} come from the first lens bin, and are due to the presence of BAO in the clustering signal. Given we are considering combinations of several different lens and source bins, this appears as several peaks. The resulting impact on reducing the suppression due to baryons is that the cross-correlation method is able to remove some of the impact at large $\ell$, reducing the suppression by around 5\% at $\ell=5000$. For the tomographic bin 3, the suppression is actually increased at large scales, due to there being more overlap with the lens samples, which is consistent with Figure~\ref{fig:krange-ratio} which shows that the smaller scales contribute more to the low $\ell$ than the standard case for this bin. This highlights the limitations of using a photometric lens sample in this method due, where it is not possible to separate sources and lenses.

Figure~\ref{fig:LUnulplots} illustrates the impact of the LUnul method method on the data vector in our fiducial setup for a representative multipole $\ell=500$. The upper left panel shows the Limber integrand for each tomographic bin pair. Curves with more support at low $k$ correspond to bin pairs at higher source redshift. Importantly, all curves asymptote to each other at very high $k$. This corresponds to the low-redshift limit, where the lensing kernel behaves as $q^i(\chi) = a \chi + \mathcal{O}(\chi^2)$ for an $i$-independent constant $a$. The lower left panel of Figure~\ref{fig:LUnulplots} shows the cumulative contribution in $k$ to the angular power spectrum of each bin pair. This shows many of the bin pairs receive significant contribution from $k > 1 \, h/{\rm Mpc}$ (black vertical dashed line) at this $\ell$.

In the upper right panel of Figure~\ref{fig:LUnulplots} we show each row of $\mathbf{L}_\ell^{-1} \mathbf{c}_\ell$, with the cumulative contribution in $k$ shown in the lower right panel. Notably, the transformed modes are now mostly all confined to lower $k$. This may seem surprising given the original data vector had significant contributions from high $k$, but may be understood by noting that at high $k$ the integrand is independent of the bin pair. This allows us to take the difference between two bin pairs and approximately cancel the contribution to the integrand from the highest $k$-modes. This is essentially what the LUnul transform is doing. This procedure leaves one mode still having significant support at the highest $k$; the transform leaves the highest-redshift bin pair spectrum unchanged.

The LUnul method has the advantage that it is constructed independently of the radial window function, so in principle can be applied to an arbitrary tracer (intrinsic alignments, photometric galaxy clustering, etc.). It has the disadvantage of having significant freedom in the numerical implementation, and, unlike the BNT transform, operates at the level of the power spectrum rather than the shear signal itself, meaning it cannot be straightforwardly applied to higher-order statistics or in the non-Limber regime.

\section{Impact of nulling strategies on cosmological constraints}
\label{sec:fisher}
\begin{figure}
    \centering
    \includegraphics[width=\columnwidth]{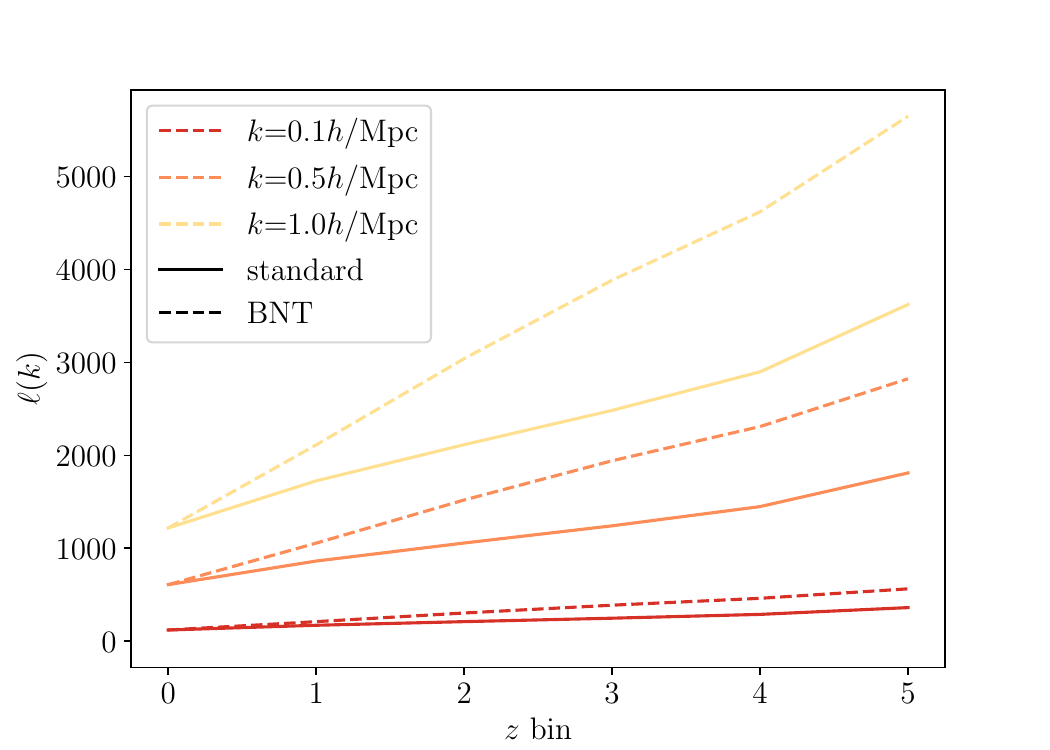}
    \caption{A cut in $\ell$ for each tomographic bin defined for a given value of $k$ using Equation~\ref{eqn:Kmax-lmax}. For the BNT case we use the BNT transformed lensing kernels shown in the lower panel of Figure~\ref{fig:kernel}. Since the BNT transformed kernels peak at higher values of $z$ compared to the untransformed case, with BNT we can go to higher $\ell$ for the same value of $k_{\rm max}$. }
    \label{fig:kcuts}
\end{figure}

Having understood how the different nulling methods impact the contribution of different $k$ ranges to the power spectrum at each $\ell$, we now consider the impact of applying scale cuts. We will attempt to remove scales beyond $k_{\rm max}$, which can be approximately related to an $\ell_{\rm max}$ for a given bin pair as,
\begin{equation}
\ell^{ij}(k) = k \chi^{ij} \, ,
\label{eqn:Kmax-lmax}
\end{equation}
where $\chi^{ij}$ is the comoving distance at the median of the lensing kernel of the lowest redshift bin in the pair $(ij)$ -- for the BNT method we use the transformed kernel. The $\ell$ cut for a given $k_{\rm max}$ is shown for each tomographic bin in Figure~\ref{fig:kcuts}. There are several alternative ways in which scale-cuts can be implemented, such as a redshift dependent $k$-cut or different $\ell$-cut per tomographic bin pair \citep{desy3/etal:2022}. As we do not seek to find optimal scale cuts in this analysis but to simply compare different methods we have only considered one approach, however this choice may impact the results reported here. 

We perform a Fisher forecast to compare the three different nulling methods to the standard cosmic shear case, for which we constrain the parameters $\{h, \Omega_{\rm b}h^2, S_8, n_{\rm s}, \Omega_{\rm m}, w_0, w_{\rm a} \}$ as well as the linear galaxy bias for the cross-correlation method. To isolate the impact of nulling on the cosmology we fix all data and astrophysical systematics parameters, like photometric redshift uncertainty, multiplicative shear bias and intrinsic alignment, to zero. We estimate the bias in cosmological parameters due to baryon feedback by assuming $\Theta_{\rm AGN}=7.8$ in the mock spectra and then not including any explicit modelling of baryon feedback. As such the `true' weak lensing shear power spectrum is given by 
\begin{equation}
    C_{\ell} = C_{\ell}^{\rm lens} + C_{\ell}^{\rm sys} \, ,
\end{equation}
where $C_{\ell}^{\rm sys}$ is the change due to baryons. The Fisher matrix~\citep{1997ApJ...480...22T} is given by
\begin{equation}
    F_{ij} = \sum_{\ell}
    \sum_{ab}\frac{dC^{a}_{\ell}}
{dp_i}
\left(\Sigma^{-1}_\ell\right)_{ab}\frac{dC^b_{\ell}}{dp_j} \, ,
\end{equation}
for parameters $p_i$ and $p_j$, where $a,b$ are vectorized bin pair indices and $\Sigma_{\ell}$ is the covariance matrix across bin pairs, which we assume receives contributions only from the Gaussian term, with partial-sky effects accounted for with a simple $f_{\rm sky}$ scaling. From the Fisher matrix the parameter covariance is then given by
\begin{equation}
    {\rm cov} [\hat{p}_i,\hat{p}_j ] = (F^{-1})_{ij} \, .
\end{equation}
The parameter bias~\citep{2008MNRAS.391..228A} is computed from the Fisher matrix and the bias vector $B_j$ as
\begin{equation}
    b[\hat{p}_i]=(F^{-1})_{ij}B_j \, ,
\end{equation}
where
\begin{equation}
    B_j = \sum_{\ell} 
    \sum_{ab}C_{\ell}^{{\rm sys},a} \left(\Sigma^{-1}_\ell\right)_{ab}\frac{dC_{\ell}^{{\rm lens},b}}{dp_j} \, .
\end{equation}

Prior to any scale cuts being applied $C_{\ell}$ is computed for an $\ell$ range $\ell_{\rm min} = 20$ to $\ell_{\rm max} = 5000$ and the derivatives are computed numerically.

\begin{figure*}[htbp]
    \centering
    \includegraphics[width=\textwidth]{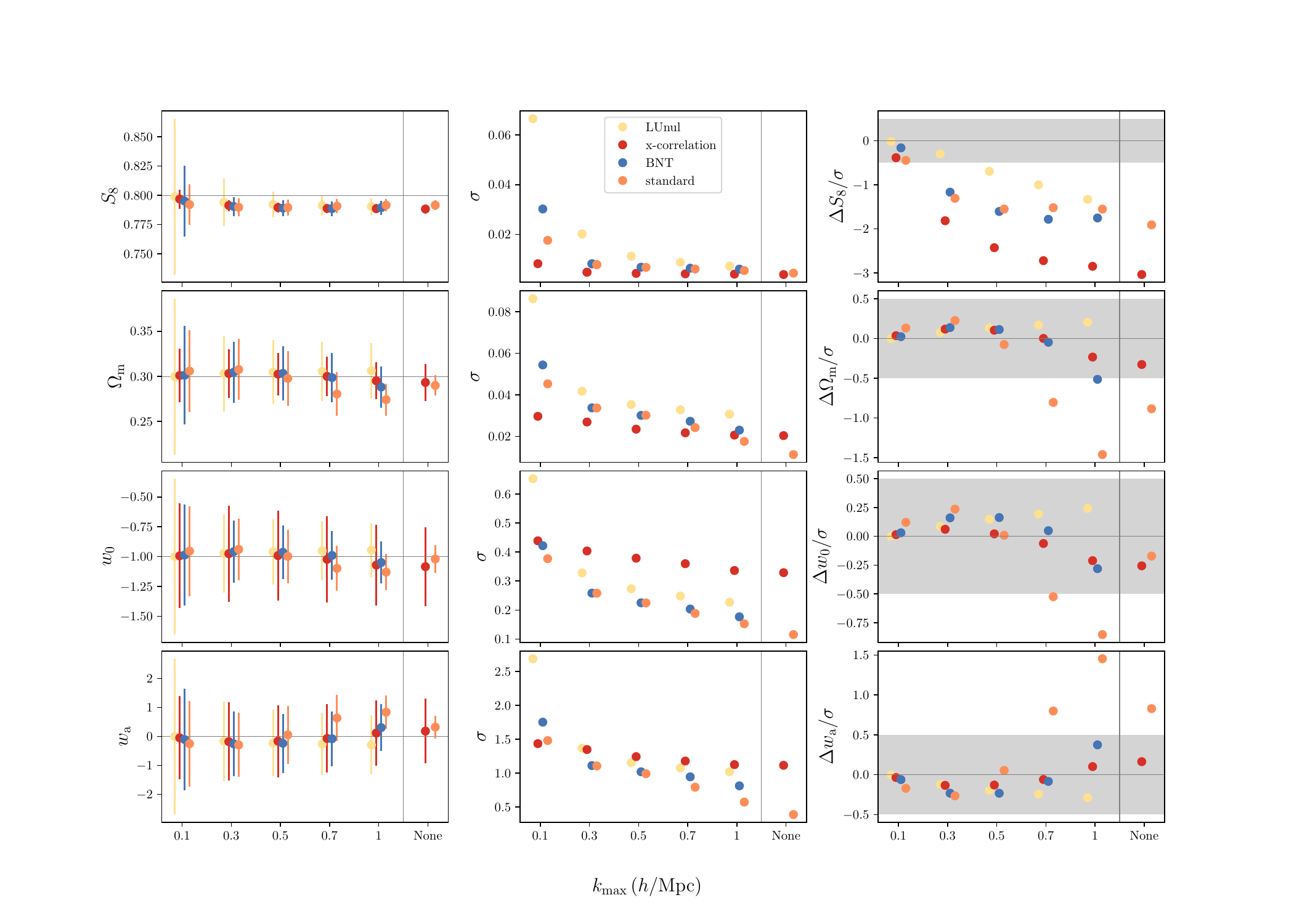}
    \caption{Constraints on parameters $\{S_8, \Omega_{\rm m}, w_0, w_{\rm a}\}$ for a range of scale cuts, defined in $k$ and transformed to $\ell$ for each tomographic bin, for the four different methods considered in this analysis: the left column shows each parameter constraint with corresponding $1\sigma$ error bars and the fiducial parameter value shown by the horizontal line, the centre column shows the error, and the right column shown the parameter bias divided by the error, with the grey region highlighting a less than $0.5 \sigma$ bias. The above plot shows how the LUnul approach consistently provides a more robust constraint across scales, whilst the BNT and standard approaches are largely consistent across a range of scale cuts. The cross-correlation method provides tighter but more biased constraints on $S_8$ and $\Omega_{\rm m}$ relative to the other methods, however is less constraining on $w_0$ and $w_a$.}
    \label{fig:tab}
\end{figure*}

\begin{figure*}[htbp]
  \centering

  \begin{subfigure}[t]{0.49\textwidth}
    \centering
    \includegraphics[width=\linewidth]{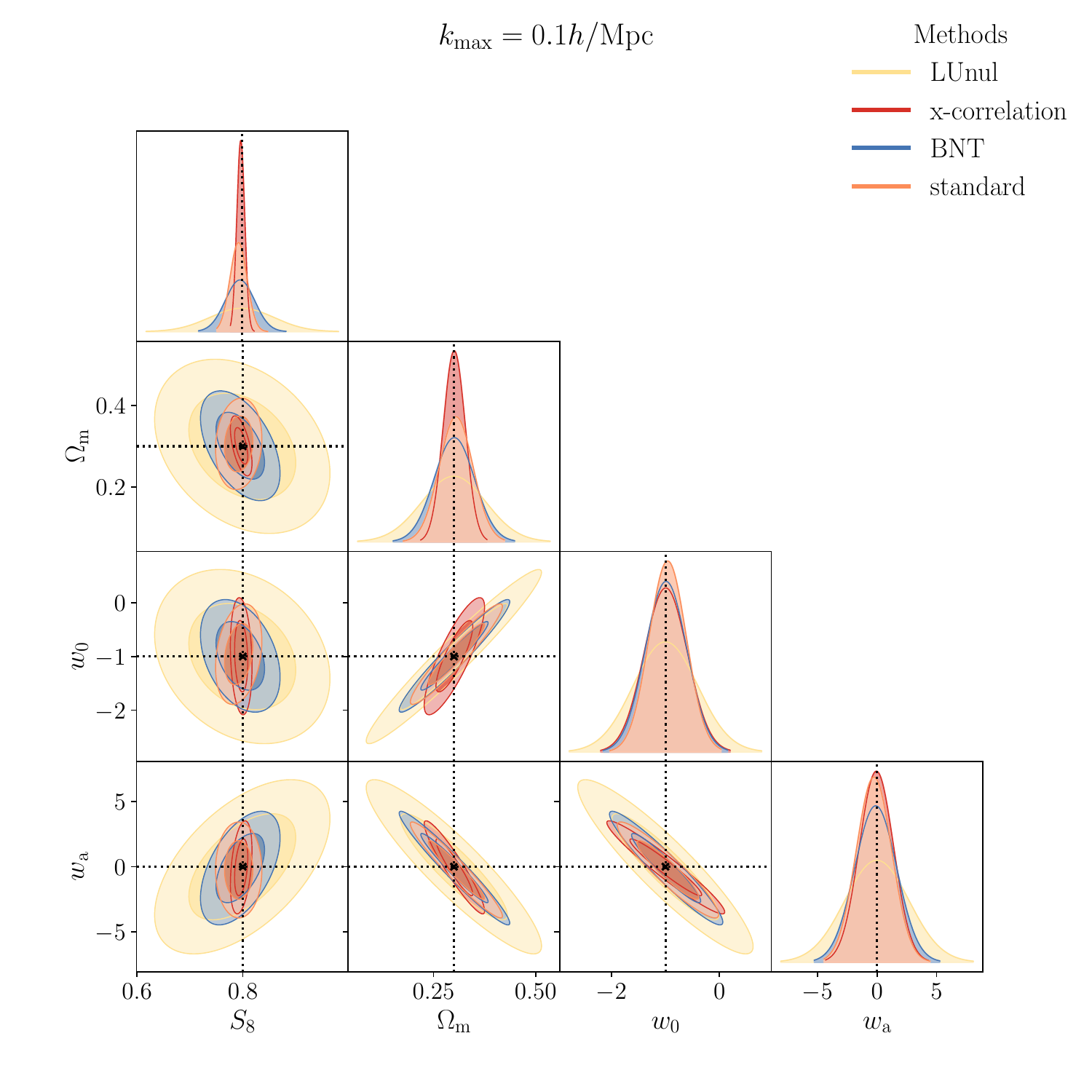}
  \end{subfigure}%
  \hfill
  \begin{subfigure}[t]{0.49\textwidth}
    \centering
    \includegraphics[width=\linewidth]{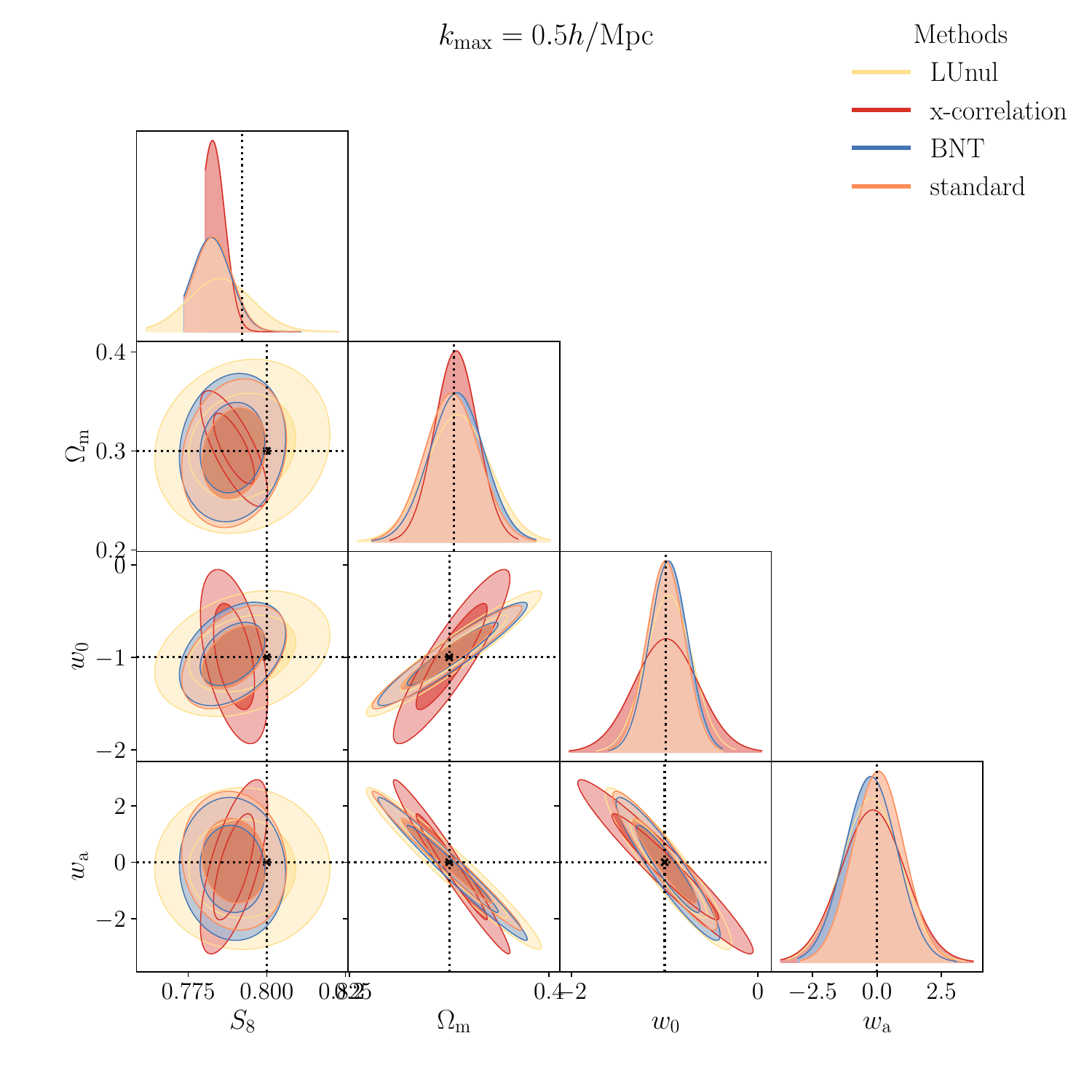}
  \end{subfigure}

  \begin{subfigure}[t]{0.49\textwidth}
    \centering
    \includegraphics[width=\linewidth]{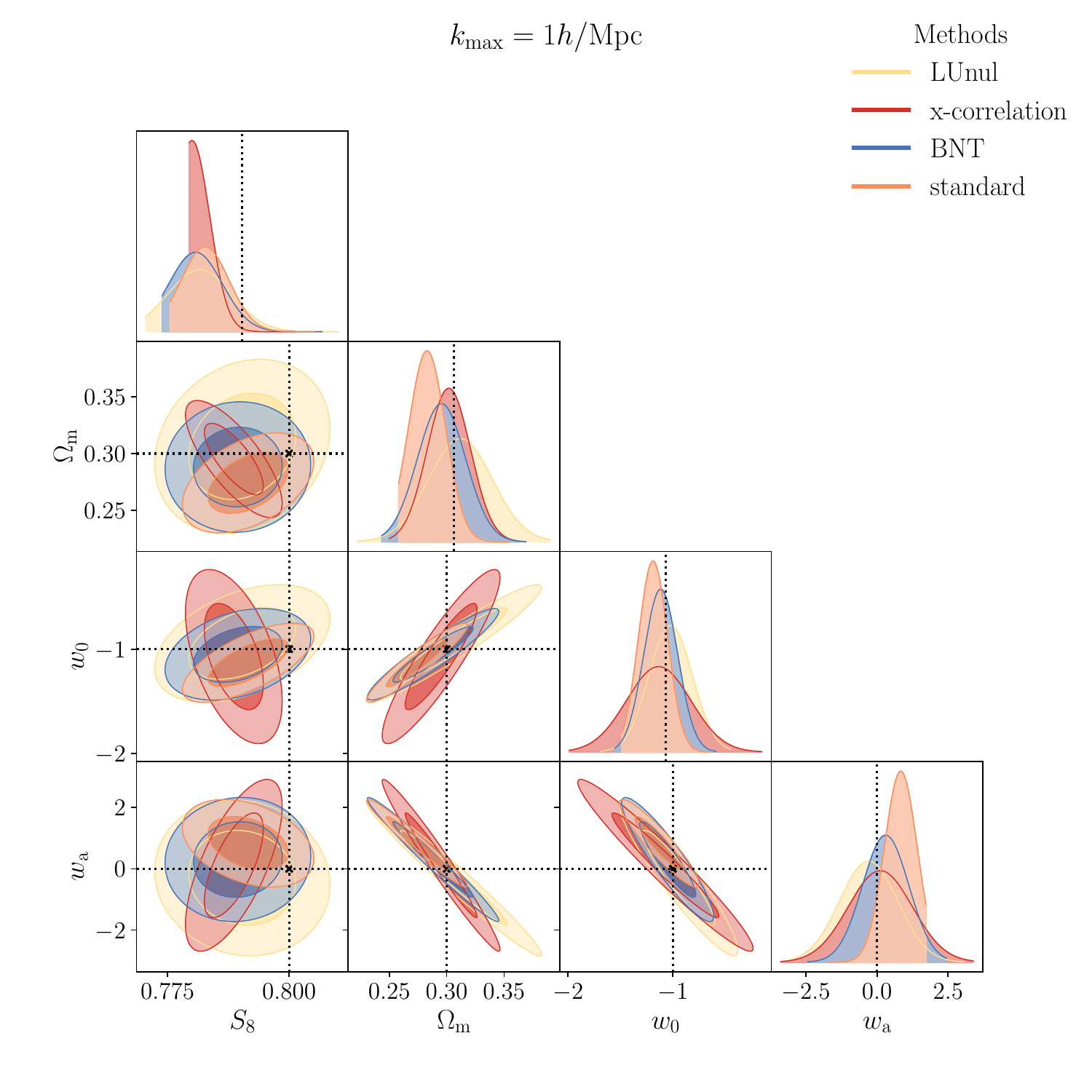}
  \end{subfigure}%
  \hfill
  \begin{subfigure}[t]{0.49\textwidth}
    \centering
    \includegraphics[width=\linewidth]{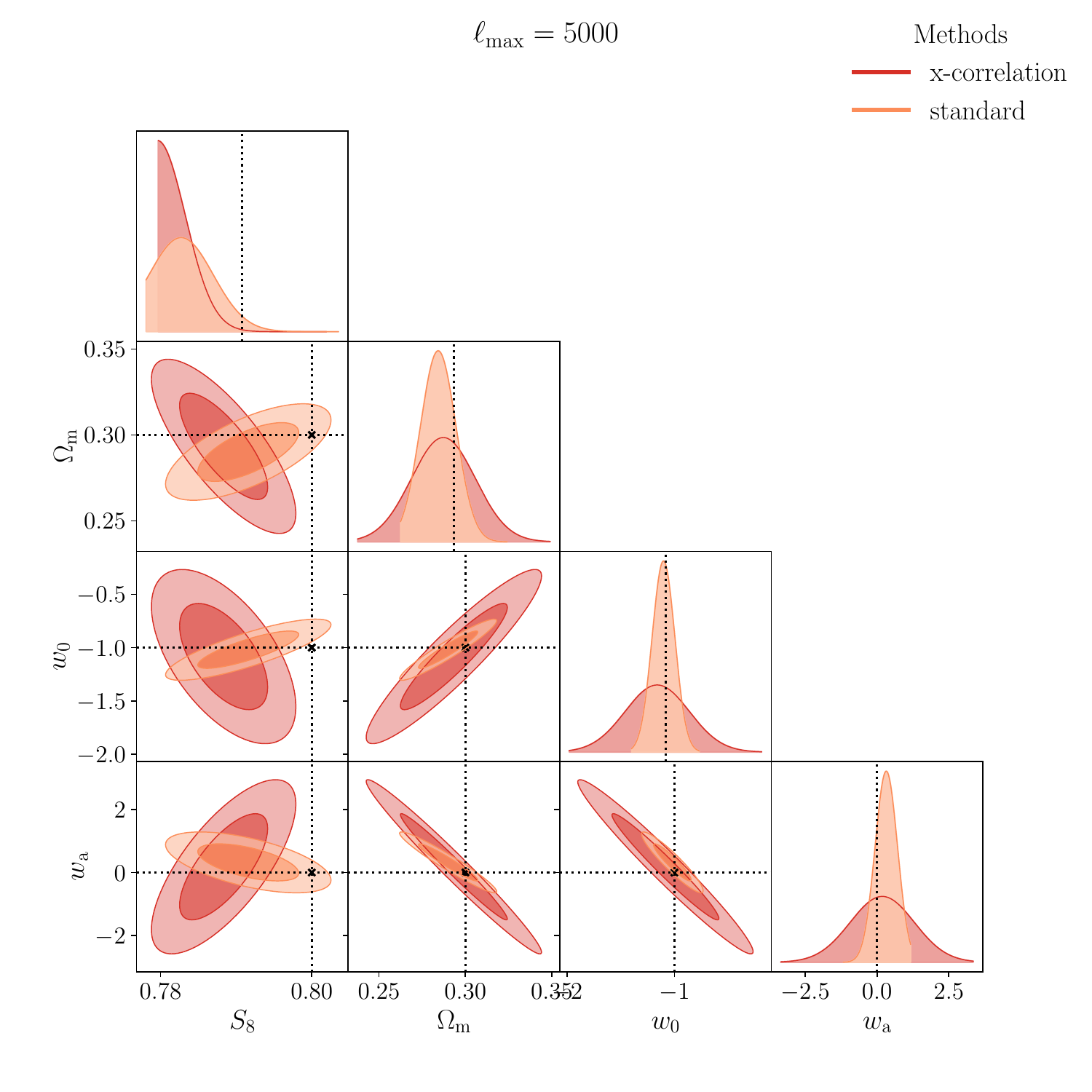}
  \end{subfigure}

  \caption{The one and two-dimensional constraints on parameters for all three nulling methods considered, compared to the standard cosmic shear, and for four example sets of scale cuts.}
  \label{fig:four-plots}
\end{figure*}

\begin{figure*}
  \begin{subfigure}[t]{0.49\textwidth}
    \centering
    \includegraphics[width=\linewidth]{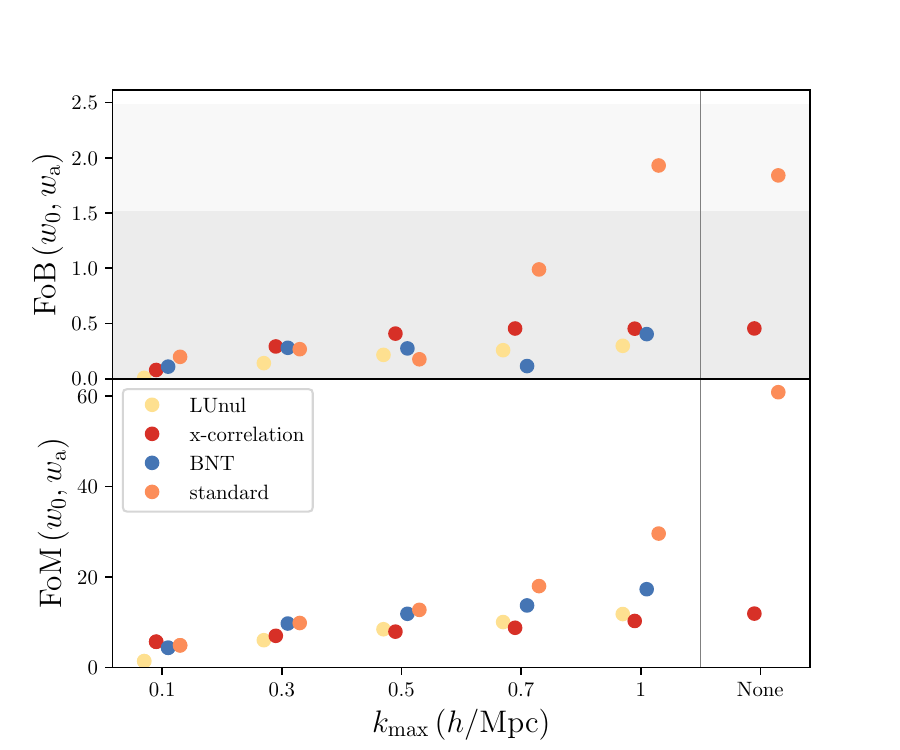}
  \end{subfigure}
  \hfill
  \begin{subfigure}[t]{0.49\textwidth}
  \centering
    \includegraphics[width=\linewidth]{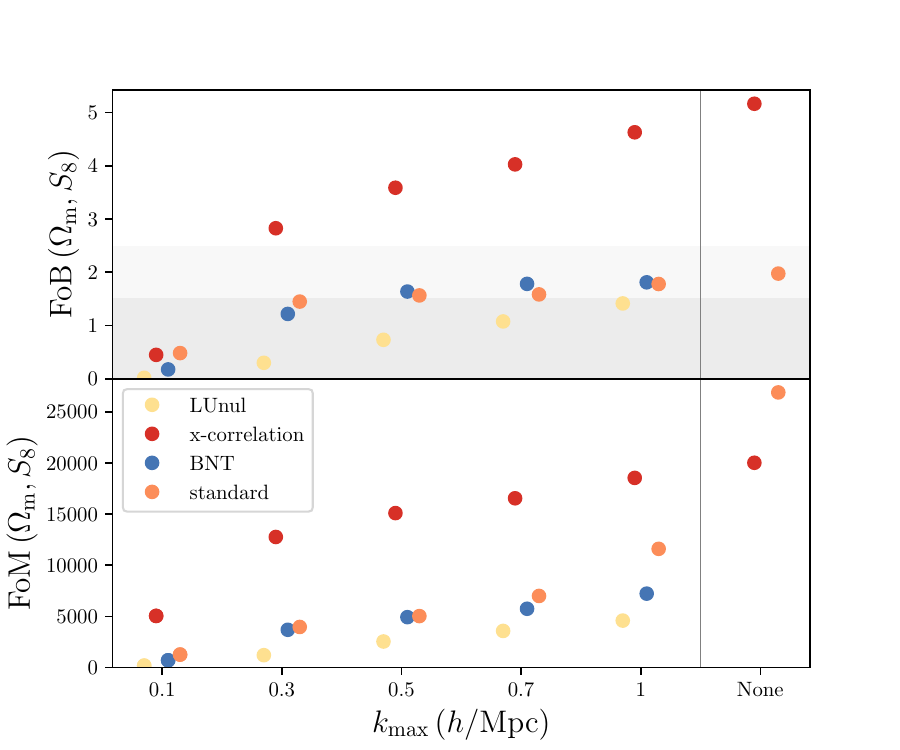}
  \end{subfigure}
  \caption{The figure-of-bias and figure-of-merit for the $(w_0,w_{\rm a})$ constraint. In the upper panel the grey regions correspond to the 1 and 2$\sigma$ level of bias in the 2D parameter space.}
    \label{fig:FoM}
\end{figure*}

Figure~\ref{fig:tab} shows a summary of the forecast constraints across a range of $k_{\rm max}$ values for the four parameters we expect to be able to constrain best with Stage-IV lensing surveys, $\{S_8,\Omega_{\rm m}, w_0, w_{\rm a}\}$. We show the corresponding $1\sigma$ error bar and the parameter bias divided by the error. The 2D parameter contours for $k_{\rm max}=(0.1,0.5,1) h/{\rm Mpc}$ and $\ell_{\rm max}=5000$ are shown in Figure~\ref{fig:four-plots}. 
We estimate the figure-of-merit (FoM) given by
\begin{equation}
    {\rm FoM}({\bf p}) = [{\rm det}(S({\bf p}))]^{-1/2}\,
\end{equation}
and the figure-of-bias (FoB) given by
\begin{equation}
    {\rm FoB}({\bf p}) = [ {\bf b}^{T} S^{-1}({\bf p})  {\bf b}]^{1/2} \, ,
\end{equation}
where $S({\bf p})$ is the covariance of parameters ${\bf p}=[p_i,p_j]$ for which we consider both the $(\Omega_{\rm m}, S_8)$ and the $(w_0,w_{\rm a})$ planes, which are shown in Figure~\ref{fig:FoM}. 

For all nulling methods, a higher $k_{\rm max}$ leads to a decrease in the error bar but also a larger bias in the parameters, as expected. Unsurprisingly, $S_8$, which approximately quantifies the amplitude of the power spectrum, is the most strongly biased parameter (due to unmodelled baryon feedback) in all cases. A bias of less than $1\sigma$ in this parameter is only achieved for the most aggressive scale cut ($k_{\rm max}=0.1 \, h/{\rm Mpc}$) in every method apart from the LUnul method; for this method, the bias in $S_8$ can be reduced to less than $0.5 \sigma$ for a less aggressive scale cut, $k_{\rm max}=0.3 \, h/{\rm Mpc}$. The cost of this, however, is that the constraining power is reduced to a similar level as the standard case at $k_{\rm max}=0.1 \, h/{\rm Mpc}$, this is mainly due to this method being a more aggressive loss of information compared to other methods. Considering constraints on $S_8$ alone, it is therefore difficult to motivate using any proposed nulling method to improve accuracy without the cost of lost precision. This statement only applies to attempts to remove sensitivity to all baryon feedback from the data vector however; if we are willing to model feedback then biases could be reduced and nulling methods may still offer significant benefits to guard against model inaccuracies on small scales. We have focused on the extreme case of ignoring baryon feedback altogether in the model, since we consider it desirable to construct a data vector entirely immune to small-scale astrophysics in order to demonstrate robustness of cosmological parameter constraints from cosmic shear. 

We find only small differences between the standard and BNT approaches, particularly in $S_8$.  Constraints on $w_0$ and $w_a$ are less biased than the standard case for higher values of $k_{\rm max}$ although the FoM remains comparable across scale-cuts, as shown in the lower panel of Figure~\ref{fig:FoM}. This is likely due to there being only six tomographic bins, as previous work \citep{gu/etal:2025} has shown that the BNT method is much more effective when there are more redshift bins, which allows for a cleaner separation of the lens redshifts that contribute to each tomographic shear signal. 

LUnul is consistently the most robust in allowing to push to smaller scales compared to the other methods. For the most stringent scale cuts the precision is significantly impacted however and the error bars become comparable with the other methods with increasing $k_{\rm max}$. This is consistent with our expectations from Figure~\ref{fig:LUnulplots} which shows the impact of the transformation on the signal and highlights how high-$k$ modes are suppressed for a given $\ell$ compared to the untransformed case. The impact of baryons here has the effect of biasing the constraint on $S_8$ high which is the opposite of effect we see in the standard cosmic shear analysis, most visible in Figure~\ref{fig:four-plots}.

The cross-correlation method provides the most biased constraints on $S_8$. As shown in Figure~\ref{fig:tab}, the parameter bias is comparable to other methods, however the error bars tighten significantly. This is likely due to the parameter information contributed by the density field. For $w_0-w_a$ we find the opposite trend, with a consistently lower FoM compared to all other methods (as shown in Figure~\ref{fig:FoM} across the set of scale cuts we consider. We also see the biggest change in degeneracies between parameters shown in Figure~\ref{fig:four-plots}, highlighting the different information content with the inclusion of clustering data.
 
For all parameters, we find that there is a nulling method which can improve parameter constraints in terms of both accuracy and precision. The BNT method is advantageous for $w_0-w_{\rm a}$, the cross-correlation approach reduces absolute biases but increases biases relative to uncertainties, and the LUnul method is successful at removing biases but its numerical implementation needs to be tuned appropriately. Our results are in agreement with previous work considering the impact of baryonic feedback on cosmic shear analyses. In all cases we recover the now well established trade-off between reducing bias in cosmological constraints versus the achievable statistical precision. We find that aggressive mitigation strategies (whether via scale cuts or nulling methods) can be effective at removing the impact of small-scale systematics like baryon feedback, but at the cost of significantly reducing constraining power. Additionally we see that the $S_8$ parameter is the most sensitive to baryonic effects, consistent with previous forecasts and survey analyses.

\section{Conclusions}
\label{sec:discussion}
In this analysis we have used a Fisher analysis to compare how nulling methods are able to reduce the impact of unmodelled baryon feedback on cosmological constraints from cosmic shear. This paper presents a simplified set of analysis choices to focus on the impact on cosmological constraints without the complication of other systematics, like the presence of intrinsic alignments, and requirements on photometric redshift and shear calibration uncertainty. Marginalising over these additional systematics will reduce constraining power and therefore our results should be understood as a comparison not taken as absolute values. Given that cosmic shear is most sensitive to the $S_8$ parameter it is not surprising that regardless of nulling choice we find that baryon feedback has the strongest impact on the $S_8$ constraint -- with similar level of bias compared to the unnulled case. We do however find that all nulling methods are able to significantly reduce the bias to less that $0.5\sigma$ shifts in $w_0-w_a$ when including small scales $(k_{\rm max}>0.7)$, compared with the unnulled case. 

LUnul is consistently the most robust method, allowing to push to smaller scales compared to the other methods, as we prioritise reducing parameter bias. Whilst this method consistently achieves a lower FoB in both $S_8-\Omega_{\rm m}$ and $w_0-w_{\rm a}$, the FoM is always lower than other methods. 
We do not find a large difference between the standard and BNT constraints, this is likely due to there being only 6 redshift bins in our fiducial setup (a realistic choice for Stage-IV surveys, e.g.~\citealt{2025arXiv250107559W}), where previous works \citep{gu/etal:2025} have forecasted that the BNT method has a bigger impact when considering a large number of well behaved redshift bins. The cross-correlation approach could potentially be set-up more optimally with a different lens selection to maximise how much of the data is used whilst also reducing the amount of overlap between lenses and sources. This approach likely works more effectively for CMB lensing since the lensing kernel peaks at around a redshift of two and thus the lens source separation is greater. Additionally we have assumed linear galaxy bias in this case, which reduces dependence on galaxy bias because of how the estimator is defined. This assumption may not be sufficient given the goal here is to push down to smaller scales. 

We have found here that the LUnul method provides consistently less biased constraints and has potential for being applied beyond a cosmic shear only analysis and with additional complexities like intrinsic alignments. The BNT method is likely more suited to the later data sets from Stage-IV surveys that will benefit from a higher number of tomographic bins. Whilst we have shown that the cross-correlation approach is able to reduce some of the suppression due to baryons, the impact is still present in larger scales due to the overlap between source and lens bins, and therefore refinement in lens selection is required. This analysis adds to the evidence that unmodelled baryon feedback leads to biased cosmological constraints and Stage-IV surveys will need to consider strong scale cuts regardless of nulling choice, or rely on accurate models of baryon feedback. These methods do however provide an alternative to the many options now available for modelling the impact of baryons. The nulling approaches consider in this work may be also useful for reducing the impact of other systematics such as intrinsic alignment and photometric redshift calibration uncertainties. 

\section*{Acknowledgements}
The authors would like to thank Peter Taylor for providing insightful feedback and comments on this draft.
The authors acknowledge support from a Royal Society University Research Fellowship. For the purpose of open access, the authors have applied a Creative Commons Attribution (CC BY) licence to any Author Accepted Manuscript version arising from this submission.

\newpage

\bibliographystyle{mnras}
\bibliography{refs} 
\appendix

\section{Additional Figures and Tables}
This appendix includes subsidiary figures which show in more details the results presented in the main part of the paper.
\begin{figure*}
    \centering
    \includegraphics[width=\textwidth]{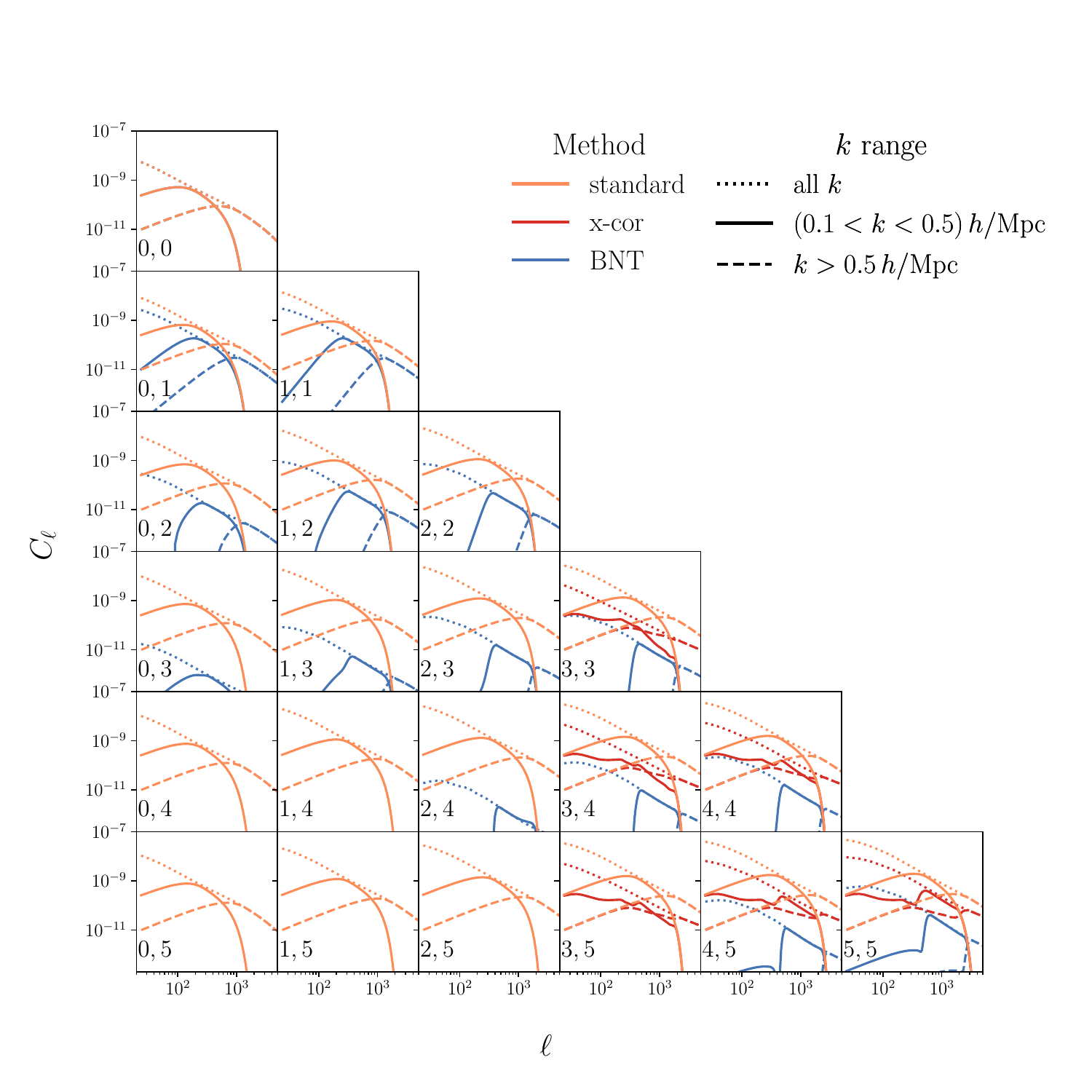}
    \caption{This figure is the power spectrum version of Figure~\ref{fig:krange-ratio}. Here we see that the high-$k$ contribution to low-$\ell$ is strongly suppressed for the BNT transformed data vector, whilst the cross-correlation only removes some of the contribution at an intermediate $\ell$ range.}
    \label{fig:krange}
\end{figure*}

\begin{table*}
\centering

\begin{tabularx}{\textwidth}{l X X X X}
\midrule
 & \textbf{Method} & \textbf{Assumptions} & \textbf{Advantages} & \textbf{Limitations} \\
\midrule

BNT &
Transform shear field to localise lensing kernel in redshift (improves $k$--$\ell$ mapping) &
Accurate redshift distributions. &
Cleaner separation of scales. &
Limited improvement with few bins. \\

\midrule

LUnul &
Apply linear transformation (e.g.\ LU decomposition) to power spectra to cancel high-$k$ contributions &
Numerical implementation choices for discretised Limber integral.  &
Generalisable beyond lensing kernels. &
Harder to extend beyond 2-point/Limber regime. \\

\midrule

Cross-correlation &
Use shear--density cross-correlations to remove low-redshift contributions &
Linear galaxy bias. &
Alternative to $3\times2$pt analyses. &
Dependent on lens selection and requires additional data and modelling. \\

\midrule
\end{tabularx}
\caption{Summary comparison of nulling methods for mitigating small-scale contributions in cosmic shear analyses.}
\label{tab:nulling_comparison}
\end{table*}

\end{document}